\documentclass[pdflatex,sns-mathphys-ay]{sn-jnl}
\usepackage{graphicx}%
\usepackage{multirow}%
\usepackage{amsmath,amssymb,amsfonts}%
\usepackage{amsthm}%
\usepackage{mathrsfs}%
\usepackage[title]{appendix}%
\usepackage{textcomp}%
\usepackage{manyfoot}%
\usepackage{booktabs}%
\usepackage{algorithm}%
\usepackage{algorithmicx}%
\usepackage{algpseudocode}%
\usepackage{listings}%
\usepackage[style=ieee]{biblatex}
\usepackage[table,xcdraw]{xcolor}%
\begin{document}

\title[Article Title]{Design and Performance of the Carruthers Geocoronal Imager}

%% \title[Article Title]{Design and Performance of the Carruthers Geocoronal Observatory}

\author*[1]{\fnm{Martin M.} \sur{Sirk}}\email{sirk@ssl.berkeley.edu}
\equalcont{These authors contributed equally to this work.}

\author[2]{\fnm{Alex M.} \sur{Zhang}}\email{alexmz2@illinois.edu}
\equalcont{These authors contributed equally to this work.}

\author[1]{\fnm{Thomas J.} \sur{Immel}}\email{immel@ssl.berkeley.edu}
\equalcont{These authors contributed equally to this work.}

\author[1]{\fnm{Jason B.} \sur{McPhate}}\email{mcphate@ssl.berkeley.edu}
\equalcont{These authors contributed equally to this work.}

\author[1]{\fnm{William W.} \sur{Craig}}\email{craig@ssl.berkeley.edu}
\equalcont{These authors contributed equally to this work.}

\author[1]{\fnm{Cathy} \sur{Chou}}\email{chouc@ssl.berkeley.edu}

\author[1]{\fnm{Kodi} \sur{Rider}}\email{rider.kodi@ssl.berkeley.edu}

\author[1]{\fnm{Edward H.} \sur{Wishnow}}\email{wishnow@ssl.berkeley.edu}

\author[1]{\fnm{Anna L.} \sur{Butterworth}}\email{butterworth@berkeley.edu}

\author[1]{\fnm{Ellen R.} \sur{Taylor}}\email{ertaylor@ssl.berkeley.edu}

\author[2]{\fnm{Lara} \sur{Waldrop}}\email{lwaldrop@illinois.edu}

\author[3]{\fnm{John T.} \sur{Clarke}}\email{jclarke@bu.edu}

%%% \author[2]{\fnm{Pratik P.} \sur{Joshi}}\email{ppjoshi2@illinois.edu}

\author[2]{\fnm{Evan M.} \sur{Widloski}}\email{evanw@evanw.org}

\author[4]{\fnm{Pascal} \sur{Blain}}\email{pascal.blain@uliege.be}

\author[4]{\fnm{J\'er\'emy} \sur{Brisbois}}\email{jbrisbois@uliege.be}

\author[4]{\fnm{Jean-Fran\c cois} \sur{Vandenrijt}}\email{jvandenrijt@uliege.be}

\affil*[1]{\orgdiv{Space Sciences Laboratory}, \orgname{University of California}, \orgaddress{\street{7 Gauss Way}, \city{Berkeley}, \postcode{94720}, \state{CA}, \country{USA}}}

\affil[2]{\orgdiv{Department of Electrical and Computer Engineering}, \orgname{University of Illinois}, \orgaddress{\street{306 N Wright St}, \city{Urbana}, \postcode{61801}, \state{IL}, \country{USA}}}

\affil[3]{\orgdiv{Department of Astronomy}, \orgname{Boston University}, \orgaddress{\street{725 Commonwealth Avenue}, \city{Boston}, \postcode{02215}, \state{MA}, \country{USA}}}

\affil[4]{\orgdiv{Centre Spatial de Liege}, \orgname{University of Liege}, \orgaddress{\street{Av. du Pré Aily}, \city{4031 Angleur}, \postcode{} \state{Wallonia}, \country{Belgium}}}


\abstract{The GeoCoronal Imager (GCI) onboard the Carruthers Geocorona Observatory is the primary scientific instrument of the mission. It is designed to measure far ultraviolet light at 121.6 nm (Lyman-$\alpha$) emitted by hydrogen (H) atoms in Earth's exosphere with the sensitivity, accuracy and precision to meet the mission's scientific objectives regarding the nature of terrestrial exospheric structure and dynamics on both global and regional scales.  The GCI is comprised of two co-aligned UV imaging systems. The Narrow Field Imager (NFI) acquires nearly continuous images of exospheric Lyman-$\alpha$ radiance near and above the Earth's limb at relatively high spatial and temporal resolution, while the Wide Field Imager (WFI) uses relatively higher optical sensitivity and a wider field of view to detect faint Lyman-$\alpha$ emission from the exosphere's outermost extent.  Both imaging channels feature identical active pixel sensor cameras, gain-intensifiers, and 6-position optical filter wheels.  This paper outlines the instrument design requirements, informed by mission science goals, as well as its performance as measured in the vacuum ultraviolet laboratory test and calibration.}
%\color{black} State a few quantitative numbers?  Resolution? Absolute Sensitivity?\color{black} }
%Design requirements outlined. Performance measured in vacuum UV at CSL.
%Spatial resolution from L1 is 2100 and 275 km for the WFI and NFI, respectively.
%The absolute radiometric throughput
%has been determined to within 20\%.
%Meets all requirements, etc.....}

\keywords{Geocorona, Far Ultraviolet, UV Imager, Lyman Alpha, Exosphere, Trappist Ales}

\maketitle

\section{Introduction}\label{intro}

The GeoCoronal Imager (GCI) is the primary scientific instrument onboard the NASA Carruthers Geocorona Observatory mission, which is designed to investigate the structure and dynamics of Earth's outermost atmospheric region known as the exosphere.  From its distant vantage near the Sun-Earth L1 Lagrange point, the GCI will obtain wide-field images of far ultraviolet (FUV) exospheric emission at 121.6 nm, the atomic hydrogen Lyman-$\alpha$ line, a bright airglow feature that is produced through resonant scattering of solar Lyman-$\alpha$ photons by the exosphere's constituent hydrogen (H) atoms.  The GCI's high resolution and high sensitivity measurements of the two-dimensional distribution of exospheric Lyman-$\alpha$ emission radiance will be inverted \cite{Widloski26,Joshi26a} (this issue) 
to reconstruct the three-dimensional distribution of the underlying exospheric H density at a temporal cadence that supports the Carruthers mission science objectives to determine the physical drivers of global exospheric structure and to characterize its periodic and transient variability.  These measurements form a solid foundation to advance understanding of fundamental exospheric physics, validate current general circulation models of the upper atmosphere and near-space plasma environment, inform magnetospheric remote sensing modalities, and improve forecasts of geomagnetic storm recovery.

Wide-field imaging of Earth's Lyman-$\alpha$ emission from a distant vantage is the ideal exospheric remote sensing approach for several reasons.  First, Lyman-$\alpha$ emission is the brightest emission line in the terrestrial airglow spectrum by far. As a result, its radiometric detection readily achieves a large signal-to-noise ratio (SNR) using integration times that easily match the temporal cadence required for the science investigation. Two dimensional imaging can detect emissions from the exosphere with the required SNR in solid angles much smaller than the hydrogen scale heights, thus also supporting the spatial resolution requirements of the science mission. Imaging supports the retrieval of volumetric H density with a plenitude of simultaneous measurements of line-of-sight (LOS) column-integrated Lyman-$\alpha$ emission radiance that span the reconstruction domain
\cite{Widloski26, Joshi26a}(this issue).  Unlike past exospheric Lyman-$\alpha$ investigations, such as the Lyman-Alpha Detectors (LADs) onboard NASA TWINS mission \cite{Nass06,Zoennchen24,Cucho-Padin18,Cucho-Padin19} or the Global UltraViolet Imager (GUVI) onboard NASA TIMED mission \cite{Christensen2003,Waldrop13,Qin16,Joshi19}, which employ mechanical scanning to obtain the viewing geometry diversity needed for H density reconstruction, two-dimensional imaging provides the needed constraints simultaneously in a single image frame.  Lyman-$\alpha$ imaging with a sufficiently large optical field-of-view (FOV) and from a sufficiently distant vantage, well outside of the exosphere itself, ensures that the image captures its outermost extent, in support of H density retrieval on fully global scales.

Owing to the vastness of the terrestrial exosphere, which extends from the exobase near $\sim$500 km altitude above Earth's surface to distances beyond 25 earth radii (Re, where 1 Re = 6371 km), along with the challenge of sensor deployment to deep space vantages, very few wide-field Lyman-$\alpha$ images of Earth's exosphere have been acquired to date.  The seminal images of Earth's global FUV airglow, acquired from the lunar surface by the Apollo 16 mission \cite{Carruthers72}, revealed the presence of significant exospheric Lyman-$\alpha$ emission beyond the edge of the camera's field-of-view ($\sim$20 Re from nadir). 
\color{black}Galileo UVS imaged an extensive Lyman-$\alpha$ geotail extending apparently past the Moon ($\sim$ 60 R$_e$) \cite{Pryor26}.  \color{black}
In 2015, the Lyman-Alpha Imaging Camera (LAICA), onboard the Japanese Proximate Object Close Flyby with Optical Navigation (PROCYON) interplanetary SmallSat mission, acquired a wide-field image of the entire geocorona from its vantage more than 2,300 Re from Earth \cite{Kameda17}.  LAICA  detected significant exospheric emission out to $\sim$37 Re from nadir, and a three-dimensional reconstruction of exospheric H density using the LAICA image exhibited enhancement in the anti-sunward direction consistent with the effect of solar radiation pressure \cite{Cucho-Padin22}.  However, the LAICA image resolution did not support H density retrieval within 6 Re of Earth, a critical region for exospheric interaction with magnetospheric plasma; moreover, the snapshot nature of the single LAICA image precluded any assessment of climatological or sporadic variability.   

The GCI is designed to overcome these historical limitations by providing global coverage of exospheric Lyman-$\alpha$ emission at high spatial resolution, high optical sensitivity, and high temporal cadence.  Using dual imaging systems with co-aligned boresights, the GCI simultaneously targets the bright, optically thick Lyman-$\alpha$ emission across Earth's limb, in support of inner exospheric H density retrieval, and the relatively dim, optically thin Lyman-$\alpha$ emission originating from the exosphere's outer region. The NASA Carruthers Geocoronal Observatory mission was selected for development in late 2020. Originally named GLIDE (Global Lyman-alpha Imagers of the Dynamic Exosphere), the mission was renamed in 2022 after the renowned space physicist George Carruthers, who pioneered the UV imaging technology used by Apollo 16 \cite{Carruthers76}.  The unique opportunity for deployment into a halo orbit (R $\sim 8 \times 10^5$ km) around the Sun-Earth L1 Lagrange point, the ideal vantage for routine exospheric imaging, was afforded by the %%% planned
launch of NASA's IMAP mission \cite{IMAP24} on September 24, 2025 with the Carruthers mission hosted by an enhanced secondary payload adapter (ESPA) ring with four available ports for mission ride-shares.
%%% Mention SWFO??
This distant vantage, $\sim$270 Re %%% 240 Re is to L1, not mean halo orbit dist.
from Earth's dayside %%%% along
near
the Sun-Earth axis, provides the GCI with continuous access to geocoronal Lyman-$\alpha$ emissions over the two-year primary mission lifetime, which is scheduled to occur shortly after the strong peak of solar cycle 25 \cite{Jouve25}.
%%% \color{black}(Is this the best Citation ?).  \color{black}
The anti-sunward, Earth nadir-staring view that the imaging payload enjoys provides protection from direct solar illumination of its apertures, an enabling characteristic for low-noise measurements and high-precision determination of absolute radiances. The off-nadir slewing capability provides access to stellar calibration sources and characterization of the background Lyman-$\alpha$ emission scattered by interplanetary hydrogen (IPH) atoms.

This report outlines the science goals of the mission and the effect they have in guiding the instrument design.  It also presents the overall performance of the two imaging telescopes as measured in the vacuum ultraviolet, as well as the optical efficiencies of the individual components.

\section{Science Objectives and Requirements}\label{sec2}
\subsection{Objectives and Science Questions}

There are two overall states of geospace that are commonly considered. One is quiet-time or quiescent, where solar radiance and solar wind inputs are steadily varying and adding relatively low amounts of energy into the geospace system, from the magnetopause to the middle atmosphere. Under these conditions, the exosphere exists under the influence of balanced inputs and outputs of energy and mass, but what is the relative importance of these drivers and processes? The second state is storm-time or disturbed, where solar radiances and solar wind inputs are much more highly variable and carry more energy into the geospace system. At these times, the energy balance of the exosphere can obviously be perturbed, but to what degree and for how long? Addressing these two questions can provide insight into the state and evolution of Earth’s exosphere. 

The first objective in support of the Carruthers mission is, therefore, to determine the drivers of quiet-time exospheric structure on regional and global scales.  As described in \cite{Waldrop26}(this issue), known drivers include thermal evaporation from the exobase, charge exchange with ambient magnetospheric H$^+$ and O$^+$ ions, and solar radiation pressure \cite{Bishop1989, Beth2016}.  Each of these drivers imposes unique spatial asymmetries on the global H density distribution, and GCI observations of the exospheric Lyman-$\alpha$ emission radiance distribution will enable the  Carruthers mission to determine their relative importance in shaping the exosphere during geomagnetically quiet conditions. \color{black}

The second objective in support of the overarching science goal of the Carruthers mission is to determine the nature and origin of transient variability in exospheric structure. To address this objective, Carruthers will characterize the spatial and time scales on which exospheric densities respond to impulsive space weather events. Such a characterization will enable Carruthers to evaluate the relative significance of storm-driven increases in thermospheric temperature and magnetospheric charge exchange reaction rates in driving observed storm-time variability.

\subsection{Science Requirements}

Carruthers science requirements are driven by the scientific need for a complete characterization of H densities around the planet on timescales that support the investigation of both quiet- and storm-time processes. The exospheric H density and its relative time-variability must be determined within defined resolution elements with sufficient accuracy to support the baseline scientific investigation. These retrieval requirements in turn engender performance requirements on the GCI regarding the temporal cadence of image acquisition, on the spatial extent and resolution of the images, and on the accuracy and precision of the measured radiances that inform the retrievals.  In practice, these imaging parameters are related to one another and must be balanced to meet the goals of the science mission.  In this section, we describe the expected nature of the target and background emisson scenes and discuss how these expectations guide GCI performance requirements, which carry margin against the scientifically limiting values and thus allow for engineering performance trades.  These top-level measurement requirements are summarized in Table \ref{tab:science_requirements} and are justified in the subsections below.

\begin{table}[h]
  \caption{GCI Observational Requirements for the two altitude regimes of the Exosphere}
    \label{tab:science_requirements}
    \centering
    \arrayrulecolor{black}
    \arrayrulewidth=0.5mm
    \begin{tabular}{>{\columncolor[gray]{0.9}}c c c}
        \hline
        \rowcolor{gray!10} % 10% grey
        Performance Parameter & Inner Exosphere & Outer Exosphere \\ 
        \hline
  %      \rowcolor{gray!10} % 10% grey
        Imaging Channel & NFI & WFI \\ 
        Altitude Range & 0-7 Re & 7-25 Re \\
        Altitude Resolution & 0.1 Re (635 km) & $\sim 1/3$ Re (2120 km) \\ 
        Azimuthal Resolution & 12 $\deg$ & 12 $\deg$  \\ 
        Wavelength of Operation & 115-130 nm &  115-130 nm \\
        Brightness Range & 2.5 - 60 kRayleigh  & 200 R - 2.5 kRayleigh  \\ 
        SNR  & $>$250 at limb  & $>$50 at 7 Re  \\ 
        Measurement Cadence & 30 min & 60 min \\ 
        Required FOV from L1 & 3.4 $\deg$ & 14.6 $\deg$ \\ 
        
        \hline
    \end{tabular}   
\end{table}

\subsubsection{Measurement Target: Exospheric H Emission at Lyman-$\alpha$}

\begin{figure}
    \centering
    \includegraphics[width=0.80\linewidth]{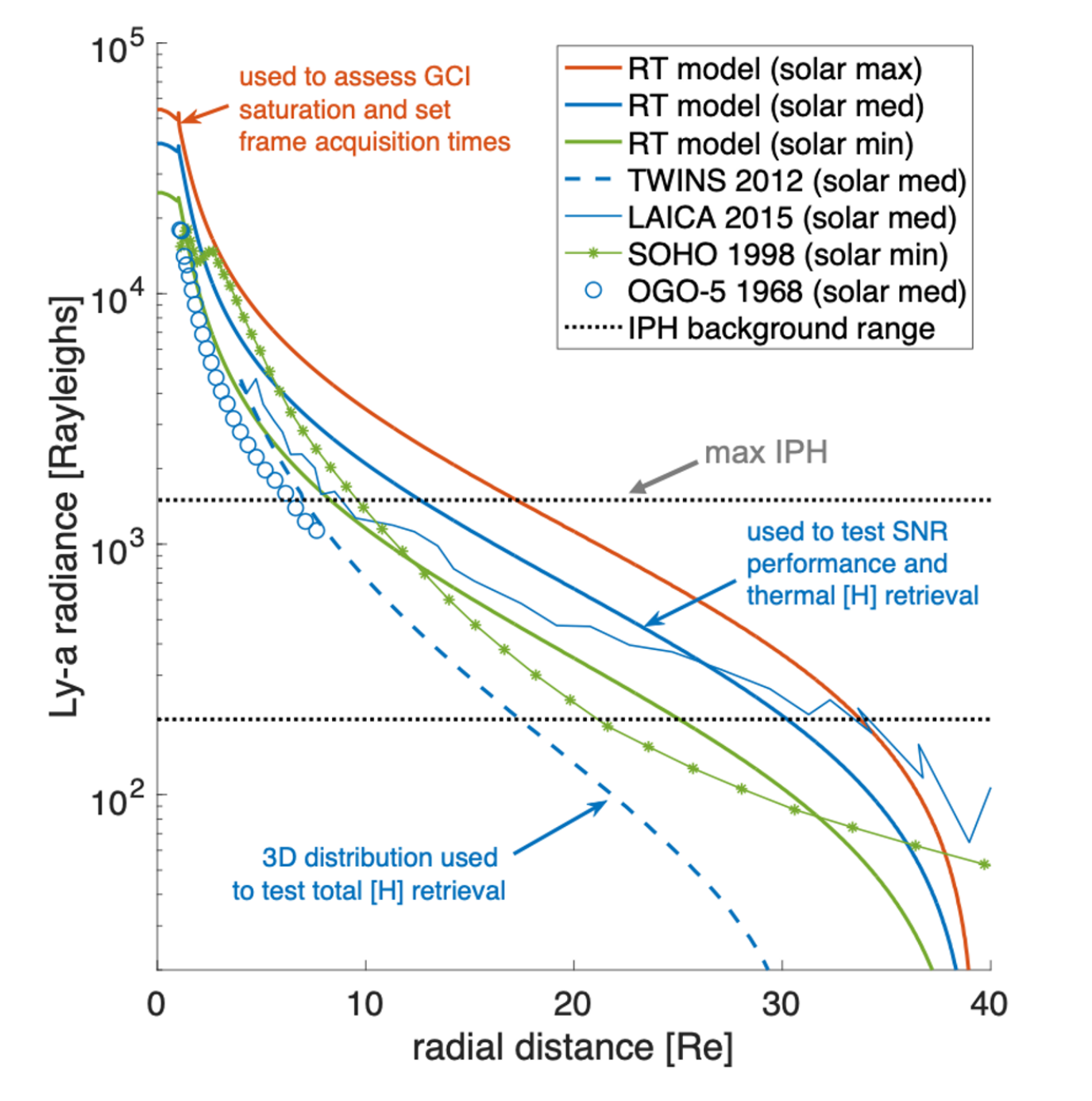}
    \caption{Exospheric Lyman-$\alpha$ radiances predicted for remote sensing views from Earth-Sun L1. Several curves show two theoretical radiance profiles and two profiles derived by analyses of Lyman-$\alpha$ measurements from TIMED/GUVI. The altitude ranges where imaging is required are shown, with the necessary imaging resolution}
    \label{fig:geocorona}
\end{figure}

The few available measurements of exospheric H Lyman-$\alpha$ (Ly-$\alpha$) radiance in the distant exosphere \cite{Zoennchen24,Kameda17,Baliukin19} are shown in Figure \ref{fig:geocorona} as a function of radial distance of the sensor line-of-sight (LOS) tangent point from earth, along with radiative-transfer (RT) model predictions for typical minimum (green), medium (blue), and maximum (red) phases of the solar cycle.    Both exospheric H density and its Ly-$\alpha$ emission radiance decreases approximately exponentially with increasing radial distance from Earth's limb, where radiance peaks sharply due to the van Rhijn effect.  As described in \cite{Waldrop26}(this issue), retrieval of exospheric H density distributions is required in the region spanning the exobase, located ~500 km altitude above Earth's surface, out to a radial distance where H density falls to 25 atoms/cm$^3$, which models indicate is within approximately 12 earth radii (Re, where 1 Re=6371 km).  This top-level retrieval requirement imposes science requirements on the GCI field-of-view, which must be sufficiently large to capture the full extent of the region of interest in a single image.   

Because the exospheric Ly-$\alpha$ radiance scale height is significantly smaller within a few Re above the limb than in the dimmer outer exospheric region of interest beyond ~5-7 Re, the required sampling of the scene needed to resolve the radial gradients in Ly-$\alpha$ radiance is itself dependent on radial distance from Earth.  Though not shown in Figure \ref{fig:geocorona}, azimuthal gradients are similarly expected to be largest in the inner exosphere, where latitudinal and diurnal variations in thermospheric temperature and vertical H flux drive corresponding variations in H transport and thus H density near the exobase \cite{Joshi22}.
%%%\color{black} (Switch to Joshi26b if Nature article is accepted).
\color{black}
These azimuthal gradients are expected to evolve with increasing radial distance, as the exosphere transitions into an approximate sun-aligned configuration due to the increasing influence of solar radiation pressure on global exospheric structure \cite{Hodges94, Hodges26}.  The resonantly scattered exospheric Ly-$\alpha$ radiance distribution also exhibits strong spherical asymmetry via its dependence on solar illumination angle, the source of the resonantly scattered geocorona emission.  For example, the volume emission rate of geocoronal Ly-$\alpha$ is significantly reduced in Earth's low altitude shadow, where direct solar Ly-$\alpha$ illumination is completely absent.  The GCI must be able to resolve expected radial and azimuthal gradients in the scene in support of retrieval of the underlying exospheric H density distribution.

Meanwhile, the exospheric Ly-$\alpha$ emission radiance scales linearly with incident solar Ly-$\alpha$ photon flux \cite{Widloski26}(this issue), which generally increases with solar activity throughout the 11-year solar cycle \cite{Emerich05}.   However, solar Ly-$\alpha$ flux can be highly time-varying on timescales of hours to days due to localized solar coronal flares and their transient evolution over a $\sim$ 28-day solar
rotation period.  Similarly, the underlying H density distribution is expected to vary both climatologically (with season and solar cycle) and sporadically in response to geomagnetic storms, particularly in the inner exosphere, where both thermal and non-thermal drivers of exospheric structure evolve on relatively short timescales (see \cite{Waldrop26} this issue for details).  The expected time-variability of the exosphere density and its scattered emission guides the required temporal cadence of density retrieval, which in turn imposes requirements on the temporal cadence of GCI image acquisition.  

The GCI is designed to meet the radial distance-dependent measurement requirements on spatial (radial and azimuthal) and temporal resolution of the geocoronal scene through the use of dual imaging systems having co-aligned boresights.  The Narrow Field Imager (NFI) is designed to target bright Ly-$\alpha$ emission from the inner exosphere at a relatively high required radial resolution of 0.1 Re and temporal resolution of 30 minutes throughout the mission.  The Wide Field Imager (WFI) is required to simultaneously image both the inner and outer exosphere regions at a lower spatial resolution of 0.33 Re and temporal resolution of 60 minutes.
\color{black} The adopted 30 and 60 minute cadences are a compromise between the desire for high temporal resolution and telemetry constraints. \color{black}
Both GCI imagers are required to achieve azimuthal sampling in the image plane of at least 12$^{\circ}$.  These measurement requirements, summarized in Table \ref{tab:science_requirements}, either meet or exceed the top-level requirements imposed on the H density retrieval itself (see \cite{Waldrop26} this issue for details).  Their specification was established through comprehensive Observing System Simulation Experiments (OSSE) \cite{OSSE26} which also accounted for the effect of measurement accuracy and precision within the individual resolution elements as described next.

\subsubsection{Measurement Accuracy and Precision}

Precision and accuracy address the repeatability of measurements and the ability to characterize systematic bias in the measurements, respectively.  Bias in the measurement of geocoronal emission radiance will arise when the target Ly-$\alpha$ signal is insufficiently isolated from background signals and/or inaccurately calibrated to physical units of photon flux.  Precision in the target radiance measurements is typically  quantified in terms of the Ly-$\alpha$ signal strength relative to the measurement noise, a quantity known as the signal-to-noise ratio (SNR).  Measurement noise is inherent to the instrument as well as the shot noise characteristics of target and background photon detection; for exospheric imaging, SNR tends to decrease with increasing radial distance from Earth, as instrument noise increasingly dominates the dim Ly-$\alpha$ scene signal.         

Systematic error in Ly-$\alpha$ radiance measurement imposes error in the retrieved H density, particularly within the optically thin region beyond ~3 Re, where the relationship between the emission radiance and the column density of the scattering H population along the viewing LOS is approximately linear \cite{Widloski26}.   Radiance measurement precision, while also affecting H density retrieval accuracy near the limb \cite{Joshi26a}, most significantly affects quantification of relative changes in H density during geomagnetic storms.  The determination of accuracy and precision of GCI Ly-$\alpha$ radiance measurement needed to meet top-level requirements on H density retrieval is based on thorough OSSE testing, in which the expected target and background scenes are specified a priori associated signal and noise levels within defined resolution elements are simulated realistically under a Monte Carlo framework \cite{Filippini26}(this issue), and H density retrieval errors are quantified for varying levels of measurement accuracy and precision.

Based on the OSSE results, the GCI is required to measure exospheric radiance at an accuracy of 30\% within the defined spatial and temporal resolution elements defined above, a requirement which maintains significant margin against the top-level requirement on H density retrieval to an accuracy of 50\%.  This requirement motivates GCI design features to support photon background characterization and/or suppression as described more below.  Meanwhile, \cite{Zhang26a, Zhang26b, Zhang26c}(this issue) detail the flowdown requirements imposed on GCI image calibration along with baseline plans for on-orbit operations and science data analysis that are designed to meet those requirements.

OSSE-based requirements on the GCI precision to meet H density retrieval requirements reflect the distinct and independent retrieval approaches at the limb \cite{Joshi26a} and in the outer exosphere \cite{Widloski26}, where SNR within the defined resolution elements is required to be $\geq$250 and $\geq$50, respectively.  A major source of measurement noise is associated with solar energetic particle (SEP) radiation in the ambient spacecraft environment, which can interact with sensor materials and manifest in the image as discrete streaks of saturated signal levels.  As a means of mitigating SEP effects on GCI image quality during a baseline 30- or 60-minute image integration time, GCI image acquisition involves the onboard co-addition of short exposure (125 msec) frames. This approach avoids SEP artifacts in the frame-stacked images at the expense of increasing frame read noise; OSSE simulations account for this engineering trade when establishing GCI precision requirements.   

\subsubsection{Background Photon Signal \#1: Interplanetary Hydrogen}

In addition to producing the exospheric Ly-$\alpha$ signal, resonant scattering of solar Ly-$\alpha$ photons also produces a significant Lyman-$\alpha$ background via its scattering by interplanetary hydrogen (IPH) atoms of interstellar origin. The potential bias introduced by its removal from exospheric-crossing lines-of-sight and the shot noise associated with this background signal must be accounted for.  

The distribution of IPH radiance across the full sky is expected to be approximately dipolar and relatively time-invariant, governed by the physics of the interaction of the heliosphere as it moves through interstellar space. The geocoronal scene (as viewed from the Earth-Sun L1 point) migrates across this structured IPH glow periodically with Earth season as well as with the spacecraft's $\sim$ 6-month halo orbit period around L1.  Spatial gradients in IPH radiance are expected to be relatively time-invariant. \color{black} Deliberate slews (performed $\sim$ weekly) of the spacecraft away from Earth nadir along the ecliptic
in advance of the Earth's position will
support systematic mapping of the IPH background. \cite{Fahr71,Yelle86}) \color{black}
%%% The subsequent removal of IPH signal from GCI images is described in \cite{Zhang26b}.  
%%The calibration requirement to characterize the IPH background in each GCI image %%without sole reliance on off-nadir IPH mapping drives the design of the GCI WFI.  %%Specifically, the WFI FOV is designed to exceed its exospheric viewing requirement, %%so that the IPH glow can be detected in the outer annulus of each WFI image, where %%it is expected to dominate measured signals beyond the likely boundary of the %%exosphere.
The estimation and removal of the background IPH radiance distribution, using both dedicated calibration data acquisition and the IPH data acquired in the outer annulus of individual WFI images, is described in detail by \cite{Zhang26b}(this issue).

\subsubsection{Background Photon Signal \#2: Terrestrial O and N$_2$ emissions}
\label{OOB}

Another potential source of uncertainty in the determination of the Lyman-$\alpha$ geocoronal brightness is the presence of thermospheric emissions at exospheric altitudes. Both imagers are responsive to terrestrial atomic oxygen emissions (OI) that are bright at both 130.4 and 135.6 nm. Though these are naturally only comparable to HI emissions at altitudes below 500 km, these bright emissions are always in the FOV of both imagers, and thus are potential sources of in-field stray light. The strategy to characterize any contribution of thermospheric emissions to the geocorona is twofold: 1) carry the capability to filter incoming light to exclude Lyman-$\alpha$ and compare the imaging to the Open (i.e. unfiltered) passband, and 2) carry the capability to implement narrow-pass filters centered at Lyman-$\alpha$ that greatly reduce contributions to the imaging signal including those originating with atomic oxygen.

The first strategy responds to the broadband responsivity of FUV-sensitive imaging systems by implementing widely used long-pass UV filters (windows) that generally show a sharp transmission cut-off at short wavelengths (115 to 130 nm), with a continuously transmissive passband out to visible wavelengths. To characterize the contribution of OI and other emissions at wavelengths longer than 121.6 nm, the imagers implement selectable UV windows that adjust the cut-off wavelength, with CaF$_2$ excluding just Lyman-$\alpha$ and then with SrF$_2$ excluding both Lyman-$\alpha$ and the OI emission at 130.4 nm. The three different resultant images allow for the characterization of emissions of OI at 135.6 nm and N$_2$ Lyman-Birge-Hopfield bands, then also the 130.4 nm altitude profile.
The second strategy implements two narrow-band interference filters to allow imaging of H Lyman-$\alpha$ in the presence of significant O and N$_2$ emissions near Earth.
The filters both have passbands at 121.6 nm (but different responsivities away from the central wavelength) which
supports the goal of obtaining H density profiles close to Earth where contaminating emissions are reduced to an allowable (and knowable) level.
This strategy requires that the peak in the instrument throughput is centered at 121 nm, with lower throughput at all longer wavelengths.
%%% With this strategy, the peak in the instrument throughput is centered at 121 nm, with lower throughput at all longer wavelengths.
Such performance cannot be achieved with a fluoride halide long-pass window (e.g. SrF$_2$), but comes from layered-deposition, narrow-pass
interference filters with peak transmission 
(albeit low throughput) at Lyman-$\alpha$. Because the focus of this measurement is the bright inner exosphere, low system efficiency ($\sim$1\%) does not impede the collection of useful emission profiles.

\begin{figure}
    \centering
    \includegraphics[width=1.0\linewidth]{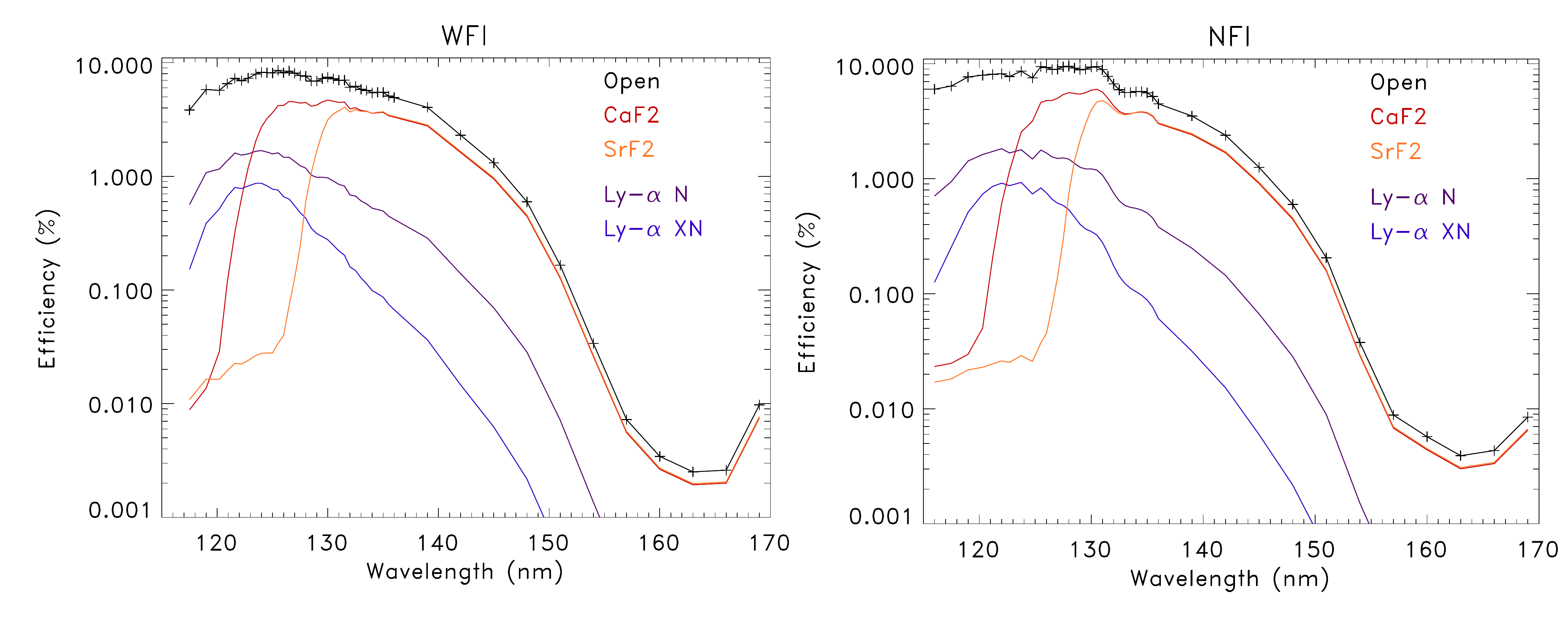}
    \caption{Comparison of Open position and the four filter efficiencies: WFI left, NFI right. Black crosses
    are the measured efficiencies for each imager for the Open position.  Colored curves are
    the product of the Open curves and the filter transmission curves provided by the manufacturers.
    The uncertainties in these curves are discussed in Section \ref{absthru}.}
    \label{fig:filter_trans}
\end{figure}

Figure \ref{fig:filter_trans} shows the measured response of each imager in the Open position (Section \ref{absthru}) as well as
the predicted response through the four filters.
The filter response curves are the product of the measured Open responses
and the filter transmissions as measured by the manufacturers ($\pm \sim $5\%).
The determination of
absolute radiometric throughput for the various filter positions and their attendant uncertainties at key wavelengths is presented in Sections \ref{absthru} and \ref{filter_trans}, and listed in Table \ref{tab:filt_eff}.

In sum, the SNR required per sample, the size of the spatial samples, and the temporal cadence of image acquisition are related parameters that are combined and balanced to meet the goals of the science mission.
In each case, the performance requirements carry margin against the scientifically limiting values, allowing for engineering performance trades. The top-level measurement requirements are shown in Table \ref{tab:science_requirements}.

\section{Payload Requirements}\label{sec3}
%%%  \color{black} Are there any pointing stability requirements? Yes, see Paper 1 \color{black}
The science requirements map to specific payload and instrument requirements. Several key requirements relate specifically to how the payload is aligned relative to the relevant components of the spacecraft, and the accuracy (offset) and precision (knowledge) of the mounting. These are not enumerated here; instead the focus is on requirements that will be tested at the instrument level.

The wider field-of-view instrument (WFI) values are in parentheses. %% ???
\textbf{}
\begin{enumerate}
    \item The NFI (WFI) shall have a Field of View $\geq$ 3.44 (14.32) deg.
    \item	The NFI (WFI) shall provide an observing mode which provides resolution (FWHM) $\leq$ 0.023 (0.080) deg.
    \item The NFI (WFI) shall provide an observing mode for which the total system efficiency at Lyman-$\alpha$ shall be better than 0.059 (0.34) count/sec/bin/kR at mission end-of-life (EOL).
    \item The NFI shall provide an observing mode for which the relative total system efficiency at 400 nm is $\leq$ 3e-5 that at Lyman-$\alpha$ over the central 0.97 deg of the FOV.
    \item The NFI(WFI) shall provide a capability to characterize out-of-band light contribution to detected signal.

\end{enumerate}

These payload requirements drive instrument design requirements (all) that will be verified in the laboratory, and mission operations requirements (\#5) that are discussed in accompanying articles \cite{Waldrop26, Craig26}(this issue).

%\subsection{Instrument Requirements}

\begin{enumerate}
    
\item The NFI (WFI) optical spot size (FWHM) shall have a diameter $<$ 40 µm (50 µm).

\item The NFI (WFI) filter wheel shall contain a narrow band Lyman-$\alpha$ filter with peak transmission $>$ 22\% and a transmission ratio at 130.4 nm not greater than 80\%.
%%% Changed < to > MMS 03/26/25

\item The NFI (WFI) shall provide an observing mode for which the relative total system efficiency at 400 nm is $\leq$ 3e-5 than at Lyman-$\alpha$ over the central 0.97 deg of the FOV.

\item The NFI (WFI) filter wheel shall contain an extra-narrow band Lyman-$\alpha$ filter with peak transmission $>$ 9\% and a relative transmission at 130.4 nm not greater than 40\%.

\item The NFI optics surface roughness and system baffling shall limit the total integrated scatter (TIS) to not more than 2\%. %% \color{black} beyond a radius  of 3.2 FWHM from optical spot center ???.\color{black}

\end{enumerate}

%The instrument will measure the emissions of H @ 121.6 nm, and O @ 130.4 and 135.6 nm with a 3-axis stabilized spacecraft,  carrying an instrument with two cameras, one for high resolution if the inner exosphere (3.5 degree field of view (FOV)) out to $\sim$7 Re tangent heights, and the other for wide field (18 degree FOV) out to ~25-30 Earth Radii (ER) with a requirement for resolution
%600 and 2100 km resolution at the mean L1 distance of $1.5 \times 10^6$ km.

\section{Carruthers Science Payload}\label{inst}
%%%%  \subsection{General Instrument Description}\label{ginst}

\begin{figure}
    \centering
    \includegraphics[width=1.0\linewidth]{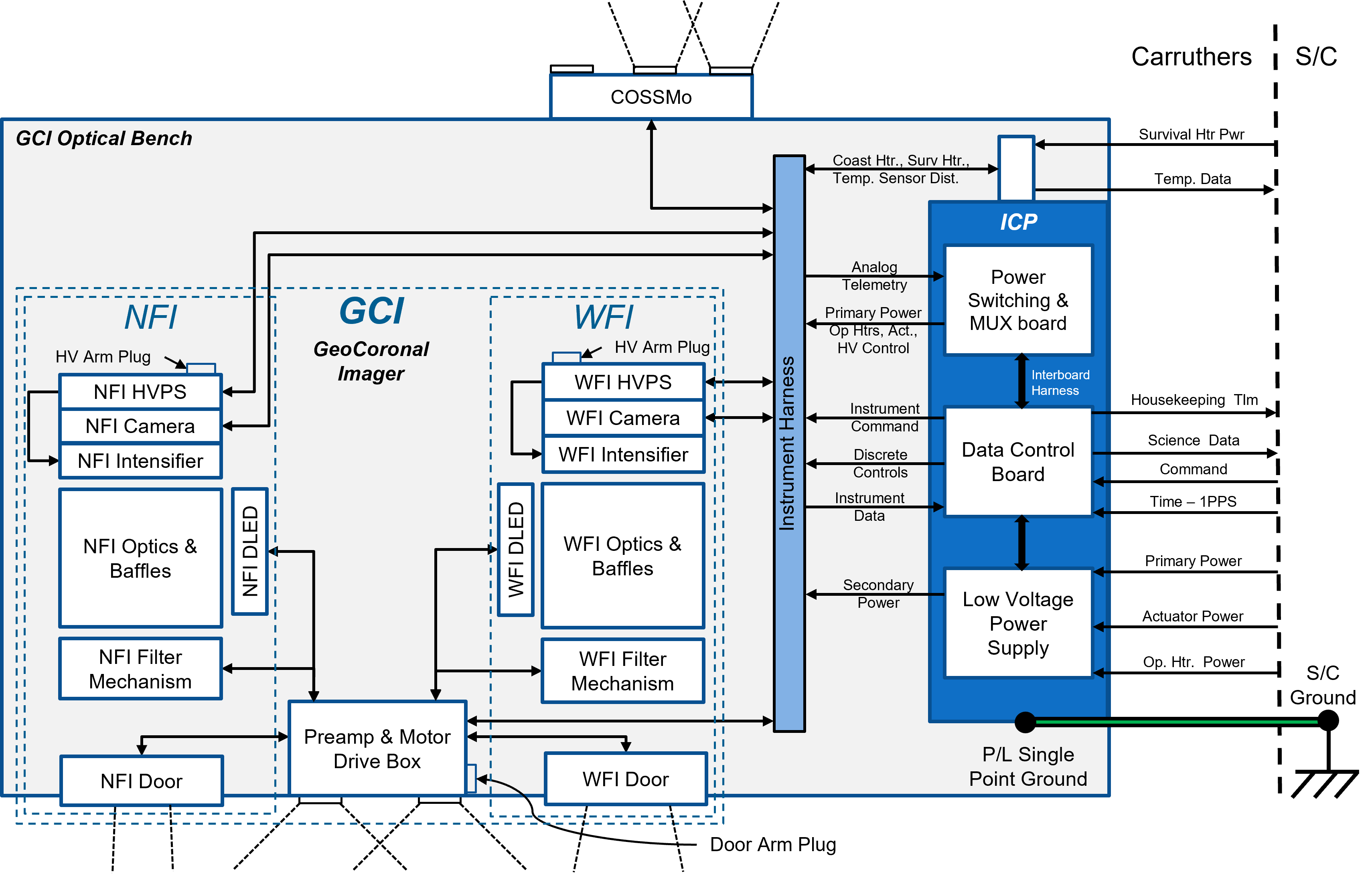}
    \caption{Simplified functional block diagram of the Carruthers payload.  Black dashed lines demarcate the FOVs of the two Imagers and the Sun Sensors (bottom), and the COSSMo experiment (top).}
    \label{fig:fbd}
\end{figure}

\begin{figure}
    \centering
    \includegraphics[width=1.0\linewidth]{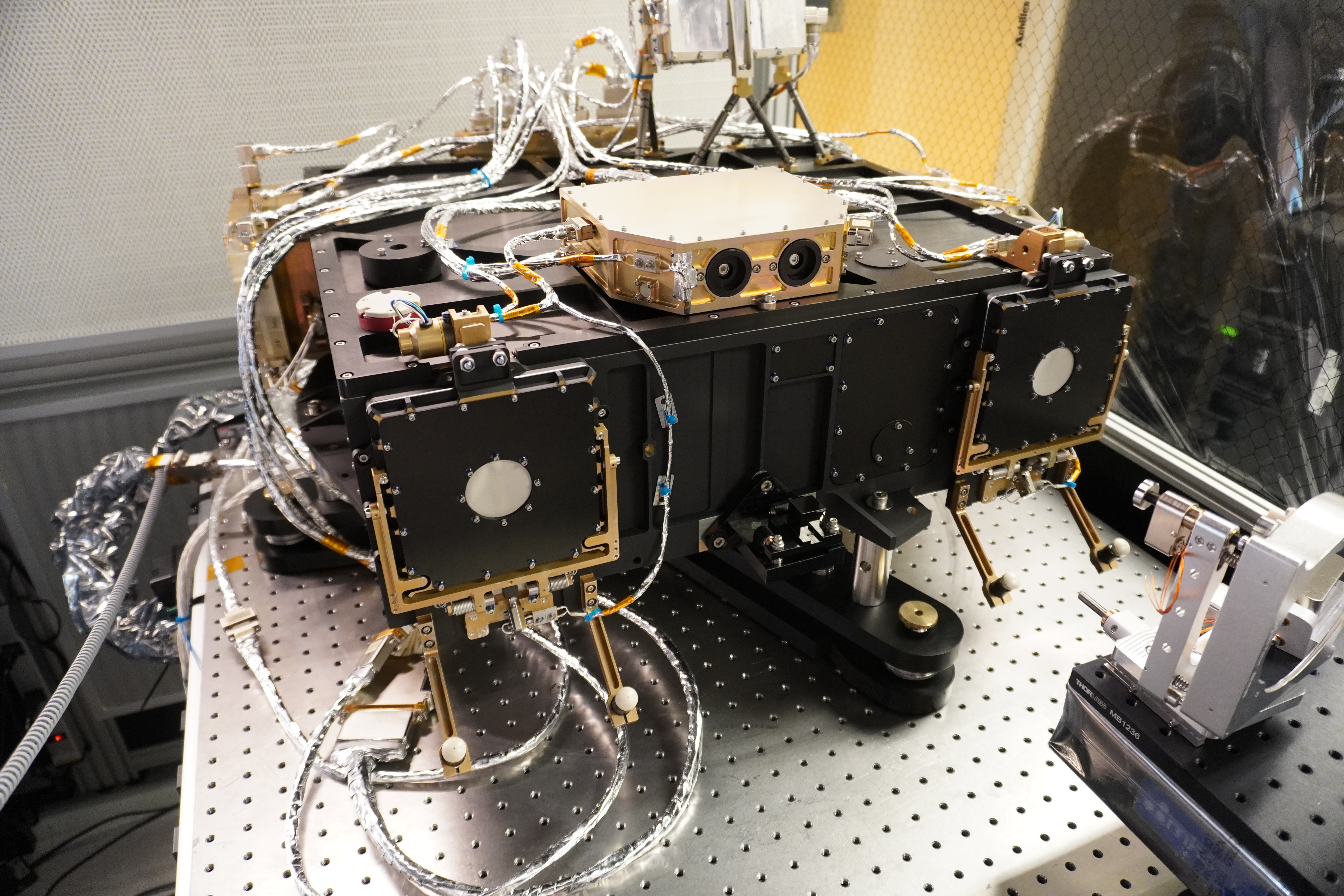}
    \caption{The Carruthers science payload: pictured are the telescope aperture one-shot door mechanisms at front in their closed positions,
    the COSSMo mounted on top (on triangular bipod struts),
    and the pair of sun sensor diodes (FOVs shown in the block diagram of Figure \ref{fig:fbd}) that are monitored by the ICP to reduce the MCP high voltages for instrument safety (visible in their own assembly atop the payload box (gold box, front, center)). Harnesses can be seen to lead back to the ICP that is on the sunward/rear (hidden) side of the payload box. The two star trackers (not yet installed in this photo) are mounted (facing roughly forward) between the two doors.}
    \label{fig:payload-photo}
\end{figure}

Figure \ref{fig:fbd} shows a simplified functional block diagram of the Carruthers payload which consists of two co-aligned imaging telescopes, two UV-light intensifying detector assemblies, two high voltage power supplies, and an instrument control package (ICP) that provides the data and power interface to the spacecraft. All of these are integrated with a single mechanical box that provides the optical bench. The walls of the box support the mounting of the cameras and ICP, as well as the lid that completes the hermetic enclosure
which provides a mounting area for a 
two channel (121.6 nm, and 0.1-7 nm) Boston University student solar monitoring experiment (COSSMo)\cite{Cossmo24,Maven15} and the sun sensors (see Figure \ref{fig:payload-photo}).
Also mounted on the Earth-facing side of the payload box (although not yet installed in Figure \ref{fig:payload-photo})
are two star trackers. They are oriented orthogonally to each other, point in the general anti-sun direction, and will provide the pointing accuracy required for science imaging.
The overall payload dimensions are $\sim$ 81 x 74 x 30 cm with a mass of 52.3 kg. 
The mechanical interface of the box to the spacecraft is through 3 bipods.
%%% The only component of the spacecraft that is mounted on the payload box are the star trackers which can NOT be seen in a photo of the payload prior to integration onto the spacecraft (Figure \ref{fig:payload-photo}).  This provides additional accuracy of the trackers relative to the science pointing of the imagers.  

\subsection{Instrument Control Package}\label{icp}
The Carruthers Instrument Control Package (ICP) provides the single-point electrical interface between the payload and the spacecraft.  The ICP receives power, instrument commanding, and timing synchronization from the spacecraft; conditions and distributes instrument power; controls instrument functions, and obtains housekeeping and science data. The ICP houses all of the payload control electronics and performs all on-board observation and science data processing (including co-adding of individual camera images to build science integrations, image compression, and packetization for transmission to the spacecraft bulk memory). The ICP provides two data streams to the spacecraft, one for science data and one for instrument state-of-health (housekeeping) data. Flight-software in the ICP provides stored command sequences, or macros, for instrument configuration changes and fault management.

\subsection{Telescopes}\label{telescopes}

% (Details of the telescope designs are discussed in an SPIE journal article \cite{Rider24}, from which we develop a summary of the two optical channels of the Carruthers payload, reported here.    We need a short description of the NFI and WFI optical designs to cover the details of what is shown in Figure \ref{fig:nfiwfiray} and captured in Table \ref{tab:telescopes}.  Are subsections needed, or just new paragraphs?)

\begin{figure}   %%%  [<placement-specifier>]
\centering
\includegraphics[width=1.0\textwidth]{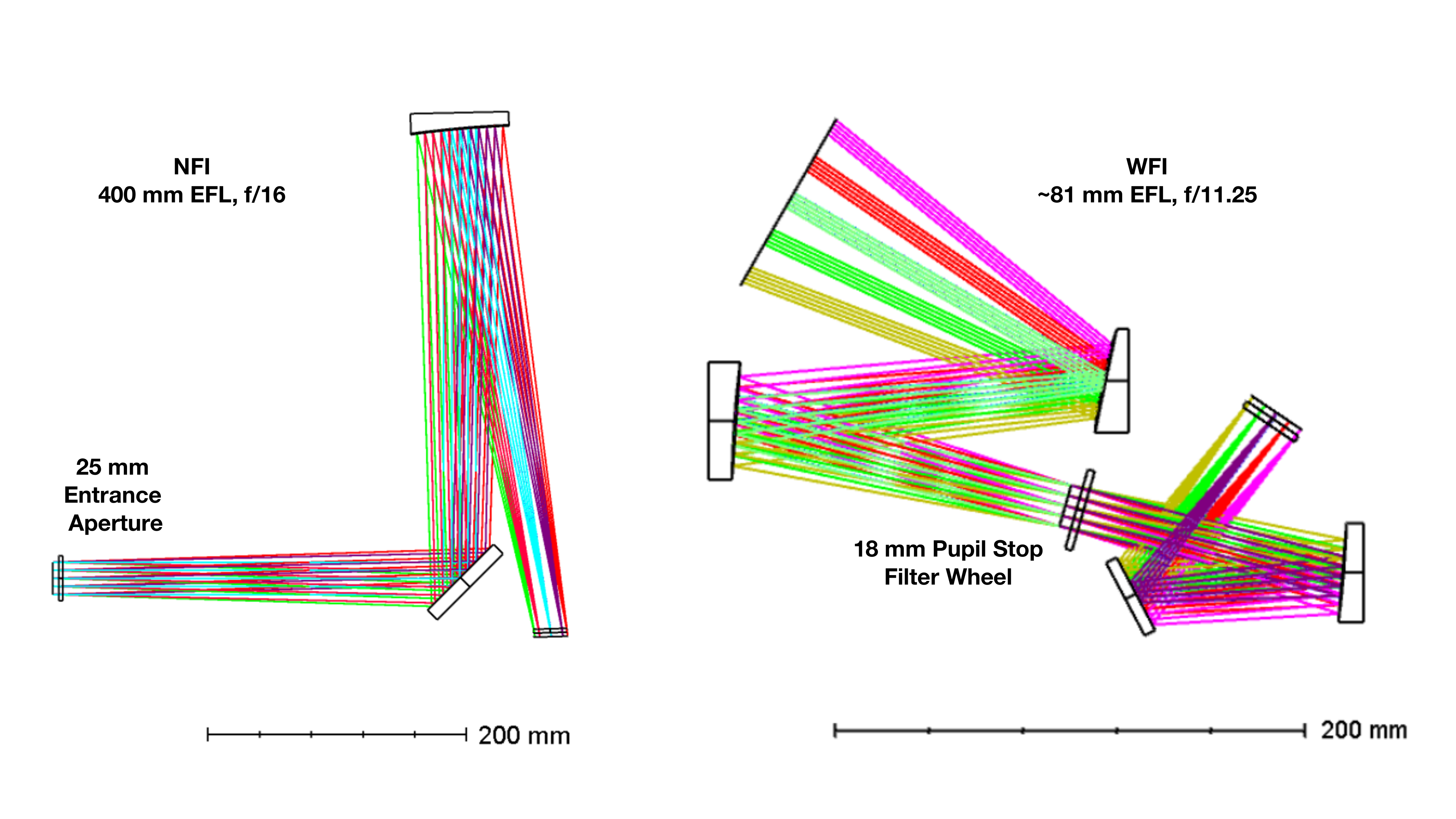}
\caption{NFI and WFI raytrace schematics showing relative mirror locations and optical paths. NFI filter wheel is mounted directly behind entrance aperture, whereas for the WFI it is mounted at an internal pupil to minimize
the angle of illumination of the filters.   WFI effective clear aperture is 7.27 mm. Adapted from \cite{Rider24}}\label{fig:nfiwfiray}
\end{figure}

\color{black}
 The design of the Carruthers payload responds to the key science objectives (Table \ref{tab:science_requirements}) to achieve imaging up to 25 Re altitudes, while simultaneously meeting requirements for 0.1 Re resolution (0.05 Re sampling) below 7 Re. The implementation of two identical cameras with different front optics is a cost-effective solution (expending engineering effort on only one camera/detector assembly design and build). 
Each telescope is constructed from fully unobstructed all-reflective off-axis anastigmatic optics 
coated with Al and over-coated with $\sim$ 26 nm MgF$_2$ (Figure \ref{fig:nfiwfiray}).
Incident radiation is focused onto two
identical Micro Channel Plate (MCP) intensified Complimentary Metal Oxide Semiconductor (CMOS) detectors.
The instrument was designed, assembled, and aligned (at visual wavelengths) at the Space Sciences Laboratory (SSL),
University of California, Berkeley \cite{Rider24}.
Thermal and UV vacuum calibrations were performed at the Centre Spatial de Li\'ege (CSL), Belgium.
The vacuum facility is described by Loicq et al., (2016)\cite{Loicq16}
with further details specific to the GCI test configuration by Rider et al., (2024)\cite{Rider24}.
The basic telescope parameters are listed in Table \ref{tab:telescopes} and
the measured imaging properties are discussed in Section \ref{calibrations}.

%%% \subsubsection{Narrow Field Imager : NFI}

The NFI imager is an off-axis Herschelian telescope -- that is, a single toroidal optical element being the only focusing optic in the system. This main optic is the second mirror in the NFI system and is located off of the axis of symmetry, with the first mirror being a flat fold mirror. The focus of any wavelength of light enables this selection of the simplest possible optic, with no need for corrective elements in a secondary mirror. The narrow field of view also supports the maintenance of good imaging from a simple toroidal mirror, where off-axis angles of only a few degrees are reached, and the packaging of this design fits well in the available space (Figure \ref{fig:nfiwfiray}).

%%% \subsubsection{Wide Field Imager : WFI}

The WFI imager requires a much larger field of view. An optical telescope that supports this large FOV while correcting for aberrations across that field is a three-mirror (one spherical, two hyperbolic aspheres), fully unobstructed, anastigmatic system \cite{Rider24}. For purposes of packaging and detector placement, an off-axis version with one fold-mirror is implemented in the WFI (Figure \ref{fig:nfiwfiray}).

\begin{table}[h]
\caption{Telescope Parameters}\label{tab:telescopes}%
\begin{tabular}{@{}cccccccc@{}}
\toprule
Instrument & Clear Aperture   & Eff Area\footnotemark[3] & Focal Length & $f$/\# & FOV\footnotemark[1] & Image Size & Res\footnotemark[2]\\
        & mm                  & cm$^2$ &mm            &     & deg & pixel & km \\
\midrule
WFI    & 7.27 &0.0295 & 81  & 11.25  & 18.1 & 512 by 512   & 2110 \\
NFI    & 25   &0.383  &400  & 16.0   &  3.5 & 1024 by 1024 & 272 \\
\botrule
\end{tabular}
%% \footnotetext{Source: This is an example of table footnote.}
\footnotetext[1]{Nominal Field of View.}
\footnotetext[2]{Resolution (FWHM) from mean L1 distance of $1.5\times 10^{6}$ km.}
\footnotetext[3]{Mean effective areas from 120 to 125 nm.}

\end{table}

\subsection{Detectors}\label{detectors}

The NFI and WFI channels employ identical cameras with UV intensifiers read out by CMOS sensors.  %%%(Table \ref{tab:mcpcmos}).  
The sealed-tube UV intensifiers, built by UC Berkeley, have 3 mm thick MgF$_2$ entrance windows and KBr opaque photocathodes (deposited on the top MCP surface), which combine to define the Open system passband of approximately 115--155 nm (MgF$_2$ is opaque below 115 nm, and the
quantum efficiency (QE) of KBr goes to zero above 155 nm \cite{Tremsin05}).
Photoelectrons from the KBr photocathode are amplified to a gain of $\sim 10^5$ by
a chevron-configuration pair of Photonis solid-edge MCPs
%%(Photonis: 33 mm diameter, solid edge, 10 µm  pore, 60:1 l/d, 13° bias angle)
(see Table \ref{tab:mcpcmos}).  The MCPs have a stack resistance of $\sim 200$ M$\Omega$ and are operated with a 2100 V bias.  The MgF$_2$ window has a 95\% transmissive layer of Ni deposited on the interior surface to prevent on-orbit build-up of charge.  The Ni layer is held at a positive potential relative to the MCP input to prevent photoelectrons from reaching the MCPs.  A 3000 V bias is used to accelerate the charge cloud from the MCPs onto a P43 phosphor faceplate at the back end of the UV intensifier which converts the electrons to a flash of green photons.
Prolonged exposure to high UV flux may ``activate" the MCP photocathodes (by raising electrons into
a metastable state) which makes them sensitive to out-of-band UV wavelengths longer than the nominal KBr cut-off at 155 nm.
To remedy this, two red LEDs ($\lambda_{peak} \sim$ 630 nm, FWHM $\sim$ 17 nm) are employed that direct light onto the MCP photocathodes
to flush the metastable states \cite{Tremsin00}.
This $\sim$ 10 minute operation to deactivate the photocathodes will be applied as needed on-orbit \cite{Waldrop26}(this issue).

The light from the phosphor screen is directed to a Fairchild CIS2521AF CMOS sensor via a reducing fiber optic taper (1.88:1) from Schott, with the 25 mm diameter active area of the MCPs falling onto the central 2048 x 2048, 6.5 µm square pixels of the sensor (full sensor size is 2560 x 2160 active pixels plus an additional 16 dark rows/columns on each side).  The combination of MCP gain, phosphor screen conversion, and sensor QE yields an average gain of about $20 \times 10^3$ electrons per detected photon distributed over approximately 12 sensor pixels, far in excess of the $\sim 8.5$ electron/pixel/frame read noise of the sensor (Table \ref{tab:mcpcmos}).
The sensors are read out by custom electronics designed and built by the Space Dynamics Laboratory (SDL) in Logan, Utah.  The readout electronics provide significant flexibility of operation of the sensors, including readout and frame rate adjustment, and on-camera pixel binning and image stacking.  The cameras are operated in a rolling shutter readout mode, with the nominal science acquisition using 125 msec frame integrations at 8 frames per second.  The NFI camera is operated with 2x2 pixel binning in the camera resulting in a 1024 x 1024 image; WFI uses 4x4 pixel binning yielding a 512 x 512 science image.  Cameras for both channels perform stacking of 480 frames, resulting in 1 minute camera integrations delivered to the Instrument Control Package (ICP)
(see Section \ref{icp}).
The ICP then performs further co-adding of the camera images to produce the ultimate science integrations
to be sent to Earth via telemetry (nominally 30 minutes for NFI and 60 minutes for WFI).
The theoretical detector performance (instrument model) is outlined %%in Appendix \ref{apendix1},
%%% by Filippini et al., (2026)
in \cite{Filippini26}(this issue).

\begin{table}[h]
\caption{Microchannel Plate and CMOS Parameters}\label{tab:mcpcmos}%
\begin{tabular}{@{}lll@{}}
\toprule
Item & Specification & Notes \\
\midrule

MCP Size      & 33 mm dia &  \\
Active Area   & 25 mm dia & 4.91 cm$^2$ total area \\
Pore Diameter & 10 $\mu$m & \\
Pore Pitch    & 12 $\mu$m  & \\
Pore Bias     & 13 deg &  \\
l/d ratio$^1$ & 60:1 & \\
Cathode       & KBr  & 40 nm thick\\
MCP Resolution & $\le 40 \mu$m  & FWHM at phosphor\\
Dark Count$^2$    & $\sim 1$ cnt/s/cm$^2$ & \\
\midrule
CMOS Size  & 16.64 x 14.04 mm & 2.336 cm$^2$ total area \\
Pixels     & 2560 x 2160 & \\
Pixel Size & 6.5 $\mu$m & square \\
Gain       & $\sim$ 17 e$^-$/DN &  \\
Read Noise & 0.5 DN/frame & 8.5 e$^-$/frame \\
Dark Current& $\sim$ 2.1 e$^{-}$/s/pixel & $\sim 5 \times 10^{6}$ e$^{-}$/s/cm$^{2}$\\
%%Low Gain   & 17 e$^-$/DN & \\
%%High Gain   & 0.5 e$^-$/DN & \\
%\DeclareGraphicsRule{}{}{}{}
\botrule
\end{tabular}
\footnotetext[1]{Pore length-to-diameter ratio}
\footnotetext[2]{One MCP photon event (count) yields $\sim$ 20,000 e$^-$ on the CMOS detector (see Section \ref{detectors}) }

\end{table}

\section{Calibrations}\label{calibrations}

\subsection{Optical Alignment}\label{subsec5.1}
To minimize the number of vacuum chamber pump-down cycles, the
GCI was first aligned in air at visual wavelengths using surrogate CMOS detectors at SSL.
%%% the Space Sciences Laboratory at the University of California Berkeley.
Each mirror and its mount and pedestal were tested as an individual sub-assembly to determine the absolute location of the mirror,
optical alignment, and wavefront error.
Once all the optics were installed, the surrogate GSE detectors were pistoned
to achieve best focus.
The visual alignment procedure is described in detail by Rider et al., (2024)\cite{Rider24}.

\subsection{Vacuum UV Ground Support Equipment}\label{subsec5.2}

The Carruthers instrument was tested in vacuum at ultraviolet wavelengths at CSL.  
The optical ground support equipment (OGSE) was inherited from that used 
to test the NASA ICON FUV instrument \cite{Frey23}.
The facilities are fully described by Loicq et al., (2016)\cite{Loicq16} with only
the most salient features reviewed here.
The vacuum test facility is comprised of three components: an UV illumination source, a collimator, and the Focal 2 chamber with hexapod, translation stage, and thermal shroud.
The illuminator is a
McPherson 225 monochromator with the following parameters: McPherson 634 Deuterium lamp, $f=1$ m with adjustable input/output slits,
rotating flat diffraction grating 1200 lines/mm, MgF$_2$ diffusers in direct contact with
pinholes (1.0, 0.3, 0.1 mm dia).
Collimator: $f=$ 1016 mm, 10 cm wide beam, 0.5 nm (RMS) mirror surface roughness,
calibrated Hamamatsu R 1081 PMT beam monitor fed by a MgF$_2$ 10\% pick-off beamsplitter.
Main Chamber: Translation stage to direct beam into either channel or onto a reflective reference cube, and a Symetrie hexapod that provides $\pm 10$ degrees in pitch and yaw motion to simulate off-axis illumination.
%% Two PMTs? Also beamsplitter - did a wavelength scan with one PMT at monitor and one PMT at GCI. ***Ratio irrelevant and will not be used in throughput section

%%This section includes details about PMT calibration that are not mentioned in Kodi's paper.
%%System setup with two PMTs.
%%PMTs also come with +- 10\% bias

%\color{black} Check where beamspliter is relative to collimator mirrors.\color{black}
The monochromator outputs a divergent beam from the selected pinhole which then encounters a beamsplitter. About 10\% of the photons are deflected into a monitoring photomultiplier tube (PMT), while the remaining photons are collimated and sent to the GCI. Thus, the irradiance at the GCI entrance apertures is known at any point in time
to within the 10\% uncertainty \cite{PTB25} in the PMT calibration.

%%The spectral scene radiance as detected by the GCI in this scenario can be modeled as $L(\lambda, i, j) = \rho \delta(\lambda - \lambda_0)\delta(i - i_0)\delta(j - j_0)$, with $\rho$ denoting the total number of photons sent to the GCI, $\lambda_0$ denoting the wavelength of the photons, and $(i_0, j_0)$ denoting the pixel index where the photons hit the GCI detector.
%%\color{black} $\delta$  is delta function \color{black}

\subsection{Impulse Response, Resolution, and Image Scale}\label{subsec5.3}

\begin{figure}   %%%  [<placement-specifier>]
\centering
\includegraphics[width=1.1\textwidth]{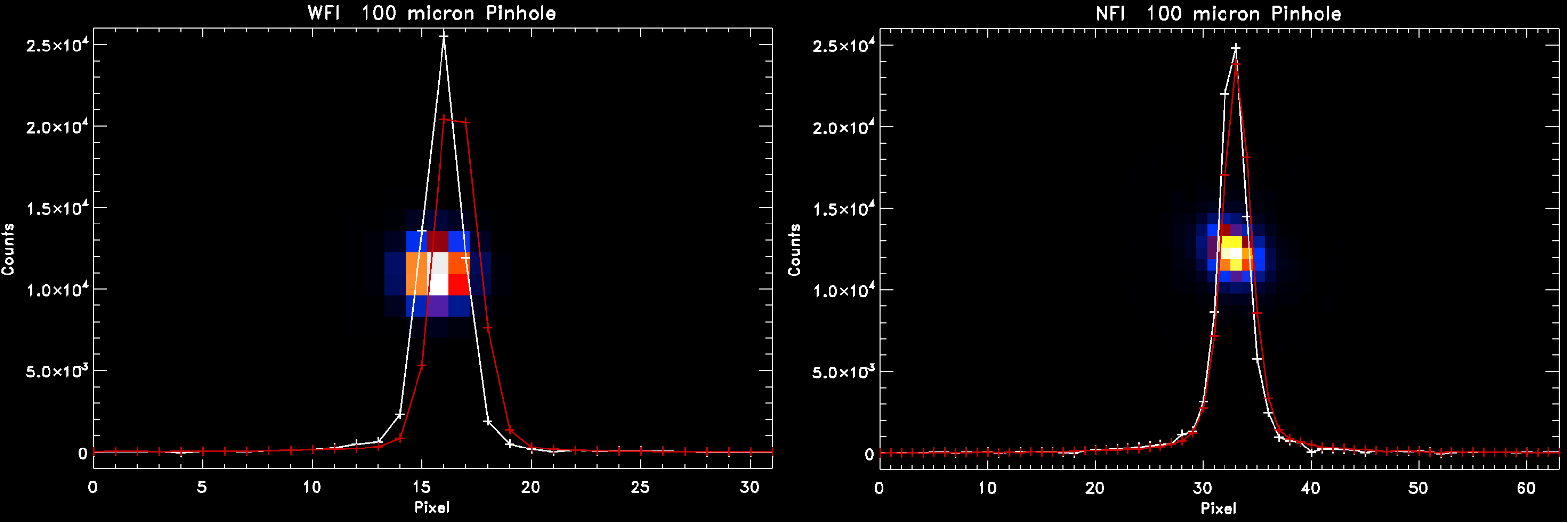}
\caption{WFI and NFI spot images and marginal distribution profiles obtained at boresight. White solid lines denote X coordinate, red denotes Y.}\label{nfiwfi}
\end{figure}

To measure the point-source spot size we employ the 0.1 mm diameter pinhole.
Short exposures were performed around the compass rose near the
edge of the detectors FOV to verify alignment and determine average image scale,
and longer exposures were performed at boresight to determine resolution.
Figure \ref{nfiwfi} shows background-subtracted images obtained with the 100 micron pinhole taken at boresight and their respective profiles in X and Y.
The measured imaging parameters are listed in Table \ref{tab:imagepars}.
The mean resolutions (FWHM) for the NFI and WFI are 0.0104 and 0.08065 degrees, respectively,
and, at a mean L1 distance of $1.5 \times 10^6$ km, corresponds to 272 and 2110 km satisfying
%% or 310 and 2390 km at mean actual distance of halo orbit 1.7x10^6 km
the imaging requirements stated in Table \ref{tab:science_requirements}.
\color{black} The off-axis images, while of shorter duration, showed comparable spot sizes, and quick-look analysis of on-orbit star fields shows good focus across the FOV for both imagers.
\color{black}

The WFI spot width at FWHM is $\approx$ 2 pixels (at 512 sampling), which means that
the scene is critically sampled (Nyquist-Shannon limited).
This is a deliberate choice in optical design to maximize the number of resolution elements in the FOV while maintaining photometric fidelity within each resolution element \cite{Stetson87}.
For the NFI the spot FWHM is $\approx$ 3 pixels (at 1024 sampling), which is an
oversampling of the scene allowing for potentially higher resolution (via deconvolution techniques \cite{Hollis92})
of the inner exosphere where gradients in radiance are large.
%%  We will achieve high resolution using deconvolution in scenes where the radiance show major gradients ...
%%%% Sounds like the Tom Leher song about Werner von Braun "A man whose allegiance   Is ruled by expedience.…"

\begin{table}[h]
\caption{Measured GCI Imaging Parameters}\label{tab:imagepars}
\begin{tabular*}{\textwidth}{@{\extracolsep\fill}llllllrr}
\toprule%
%%& \multicolumn{3}{@{}c@{}}{Element 1\footnotemark[1]} & \multicolumn{3}{@{}c@{}}{Element 2\footnotemark[2]} \\\cmidrule{2-4}\cmidrule{5-7}%
Inst & $^1$Scale & FWHM X & FWHM Y & $^2$EE 95\% & $^2$EE 98\% & FOV X & FOV Y \\
\midrule
WFI & 0.0365   & 0.0724 & 0.0889 & 0.117  & 0.189  & $>$17.84 & 18.11\\
NFI & 0.00347  & 0.00965 & 0.0111 & 0.0256 & 0.0339 & 3.51   & 3.53\\
\botrule
\end{tabular*}
\footnotetext[1]{Mean Image Scale is degrees/pixel. All other values are in degrees.}
\footnotetext[2]{Encircled Energy Radius}

\end{table}

\begin{figure}   %%%  [<placement-specifier>]\centering
\includegraphics[width=1.05\textwidth]{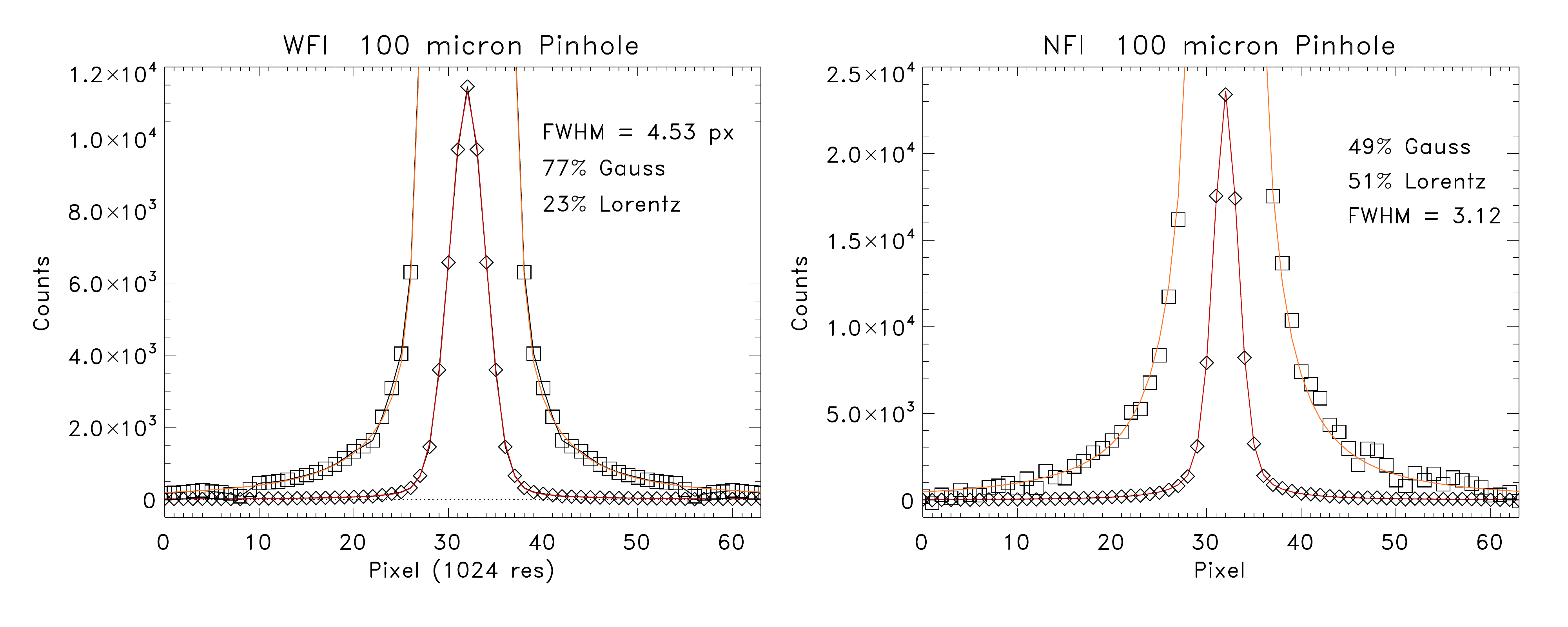}\caption{
WFI (left) and NFI (right) PSFs (diamonds) and 20x vertical enlargement (squares).
Over-plotted in both panels (color) is the least-squares fit of a Gauss plus Lorentz function model with the same FWHM.  The WFI PSF in this plot is created by resampling the original data to 1024 resolution, and,
because of asymmetries in spot shape (see Figure \ref{nfiwfi}),
the X and Y profiles are combined into a single profile.
Note that the WFI shows a broader FWHM and a greater Gaussian component than the NFI, but with similarly shaped Lorentzian wings.}\label{scatter}
\end{figure}

The instrument impulse response (or Point Source response Function (PSF))
measured at CSL
is the convolution of the optical spot, the MCP detector (tube blur),
and scattering from the telescope mirror surfaces (on orbit there will also be pointing-jitter).
The OGSE contributes additional blurring caused by scattering from the two 
collimator mirrors.
%% Diffraction and scattering by the 50 $\mu$m thick, 0.1 mm diameter foil pinhole edge is irevelelant since large-angle rays simply miss the collimator primary mirror.
The measured single mirror RMS surface roughnesses $RMS_{SR}$ are: WFI $\approx 0.25$ nm, NFI $\approx 0.32$ nm, CSL Collimator $\sim 0.5$ nm.  The worst-case predicted Total Integrated Scatter (TIS)
$ TIS = (4\pi RMS_{SR} /\lambda)^2$ is 0.78\%,
which is small compared to the observed Lorentzian wings (indicating that the observed wings are more likely caused by optical aberrations and not mirror surface scattering).
The actual PSF will be determined on-orbit by multiple observations of stars \cite{Waldrop26,Zhang26c}(this issue).
Figure \ref{scatter} shows the NFI and WFI spot profiles and least-squares fits
of a Gauss plus Lorentz function; the 20x vertical enlargement clearly shows the Lorentzian wings in the PSF in both channels that extend to about 25 pixels from the spot center (at 1024 sampling).
The 98\% Encircled Energy radius is 3.27 and 2.43
times the mean resolutions for NFI and WFI, respectively.

\subsection{Optical Distortion}\label{subsec3.3}

\begin{figure}   %%%  [<placement-specifier>]
\centering
\includegraphics[width=1.0\textwidth]{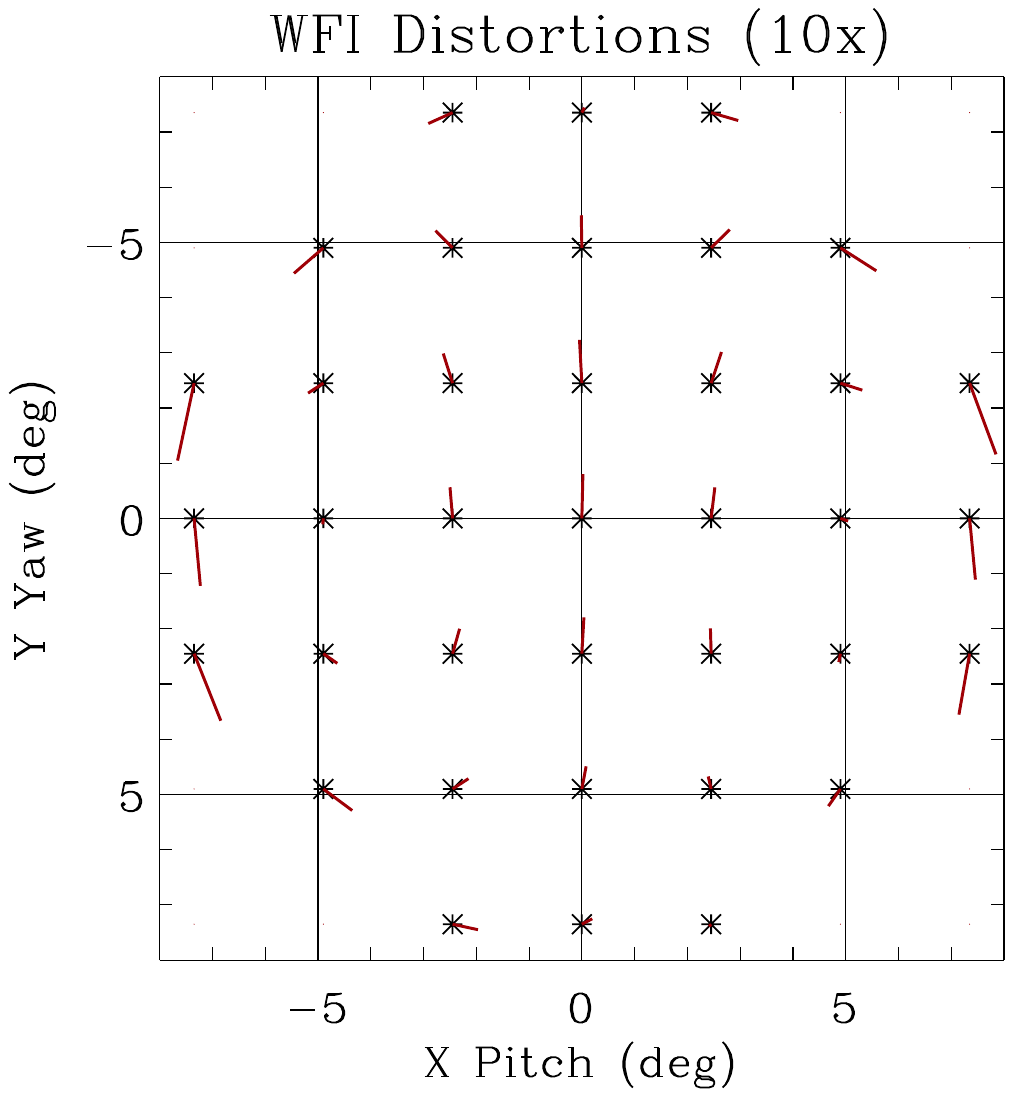}
\caption{Distortion Vector Field for the WFI.  Black asterisks are the uniform Pitch and Yaw grid fiducials spaced at 2.45 degrees. Red lines indicate direction and magnitude of distortions (times 10).}\label{distortion}
\end{figure}

\begin{figure}   %%%  [<placement-specifier>]
\centering
\includegraphics[width=1.0\textwidth]{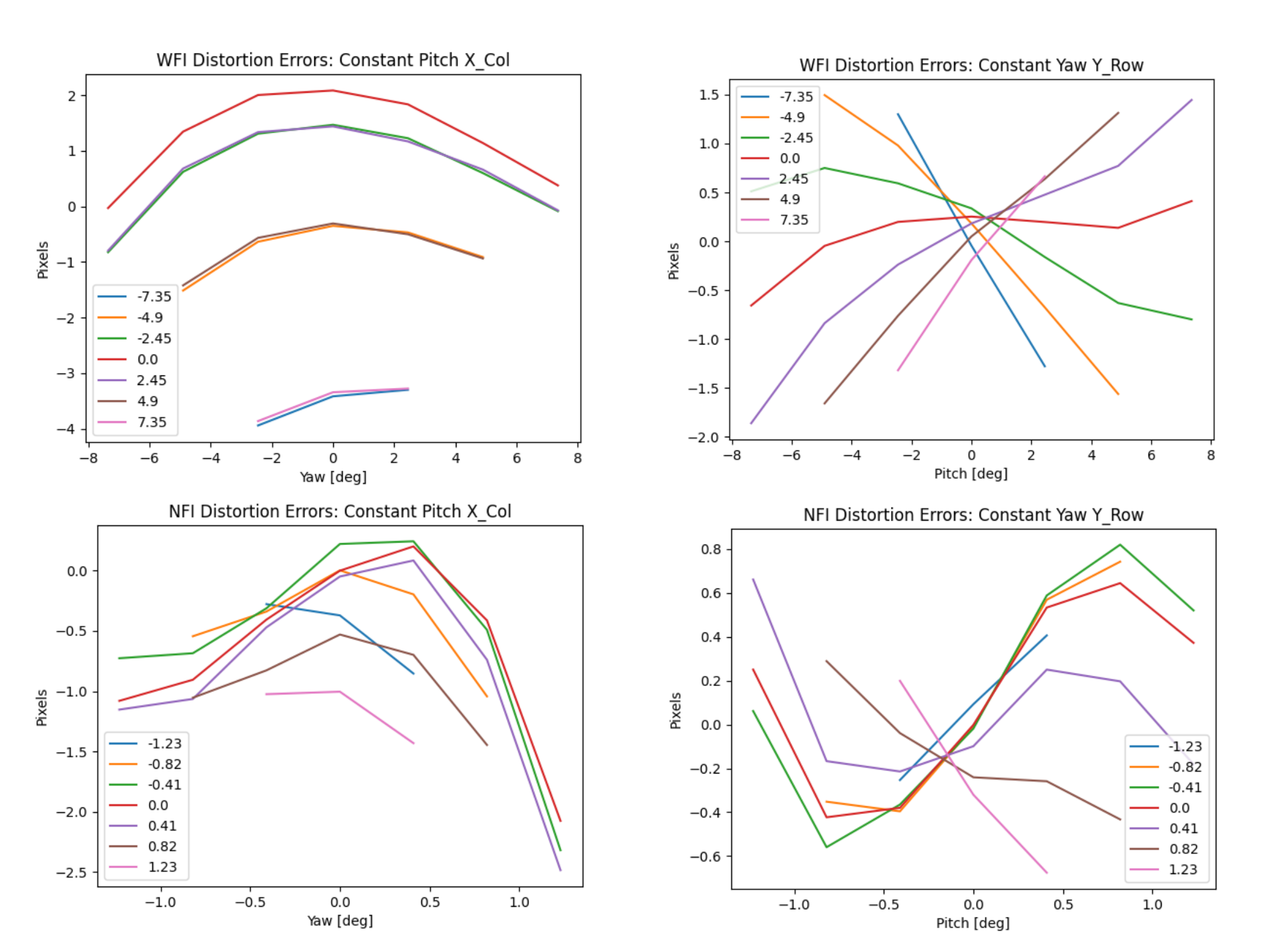}
\caption{Distortions in X and Y (pixels) for WFI and NFI as a function of off-axis angle (in degrees, insets) in each coordinate.}\label{distortion_xy}
\end{figure}

To provide good imaging over the large FOV of the WFI, a three-mirror off-axis
anastigmat is employed (see Section \ref{telescopes}).  Intrinsic to this design are optical distortions
that slightly change the image scale across the FOV in both dimensions.
The net effect is that equal areas of the sky do not map to equal areas of the detector.
Distortion maps were created by slewing the instrument in Pitch and Yaw
and obtaining images on a uniform angular grid.  Differences between the observed
image centroids and their expected positions (based on the known off-axis
grid angles and mean image scales) yield a distortion vector field.
Figure \ref{distortion} shows the distortion field for the WFI.
Third-order bivariate polynomials of the general form shown in equations (\ref{eq:dist_x_poly_sum}) and (\ref{eq:dist_y_poly_sum}) are fit to the differences via least squares to best determine the coefficients $a_{j,i}$ and $b_{j,i}$.
Because the distortions are small,
and the rotation angle between the instrument and GSE rotation stage is zero,
higher-order cross-terms were found to be negligible.
Equations (\ref{eq:dist_x_poly_expand}) and (\ref{eq:dist_y_poly_expand}) are our simplified distortion polynomials.
The original image pixel-center coordinates $(x,y)$ are warped by the polynomials
and then the Z value (counts) evaluated (by spline interpolation) onto a uniform output grid.
Each output pixel represents an equal solid angle patch of the sky,
and is effectively the number of observed counts weight-averaged over 4 original pixels.
Figure \ref{distortion_xy} shows the distortions in X and Y for both the WFI and NFI imagers. Because of the simpler optical scheme and longer focal ratio employed by the NFI, the distortions are about a factor of 2 smaller (as shown in Figure \ref{distortion_xy}). 

\begin{equation}\label{eq:dist_x_poly_sum}
X^{'} = \sum_{j=0}^{N} \sum_{i=0}^{j} a_{j,i}\, x^{j-i} y^{i}
\end{equation}

\begin{equation}\label{eq:dist_y_poly_sum}
Y^{'} = \sum_{j=0}^{N} \sum_{i=0}^{j} b_{j,i}\, y^{j-i} x^{i}
\end{equation}

\begin{equation}\label{eq:dist_x_poly_expand}
X^{'} = a_{00} + a_{10}x + a_{11}y + a_{20}x^{2} + a_{21}xy + a_{22}y^{2} + a_{30}x^{3} + a_{31}x^{2}y + a_{32}xy^{2} + a_{33}y^{3}
\end{equation}
\begin{equation}\label{eq:dist_y_poly_expand}
Y^{'} = b_{00} + b_{10}y + b_{11}x + b_{20}y^{2} + b_{21}yx + b_{22}x^{2} + b_{30}y^{3} + b_{31}y^{2}x + b_{32}yx^{2} + b_{33}x^{3}
\end{equation}

\color{black}The distortion coefficients will be re-evaluated on-orbit by imaging stars. The anti-Sun location along the ecliptic crosses the galactic plane twice a year (in June and December) which conveniently provides the CGI fields rich in stellar targets. 
\color{black}

\subsection{Absolute Radiometric Throughput}\label{absthru}

The absolute radiometric throughput needs to be well characterized on the ground. This was done in the Focal 2 vacuum chamber at CSL in the UV.
Wavelength sweeps were performed using the Open filter from 115 nm to 169 nm (with fine wavelength sampling of 0.5 nm steps around Lyman-$\alpha$ and a coarser wavelength sampling of 3.0 nm at the tails) to obtain absolute radiometric throughput $\epsilon(\lambda)$ in units of detected events / photon. The data acquired are then used to calculate the instrument QE $\eta(\lambda)$ as discussed %%%%  by Filippini et al., (2026) 
in \cite{Filippini26}(this issue).

%%%using Equation \ref{eq:opt_efficiency}.

To characterize efficiency, we need both the input photon rate at the telescope aperture and the detected event rate from the detector. The detected event rate for each wavelength is acquired by integrating the spot in the image and normalizing by the integration time of the image. Detector bias and dark counts are removed through the measurement and subtraction of dark images taken before and after each set of spot images, using the same camera settings.

\begin{figure}   %%%  [<placement-specifier>]
\centering
\includegraphics[width=1.0\textwidth]{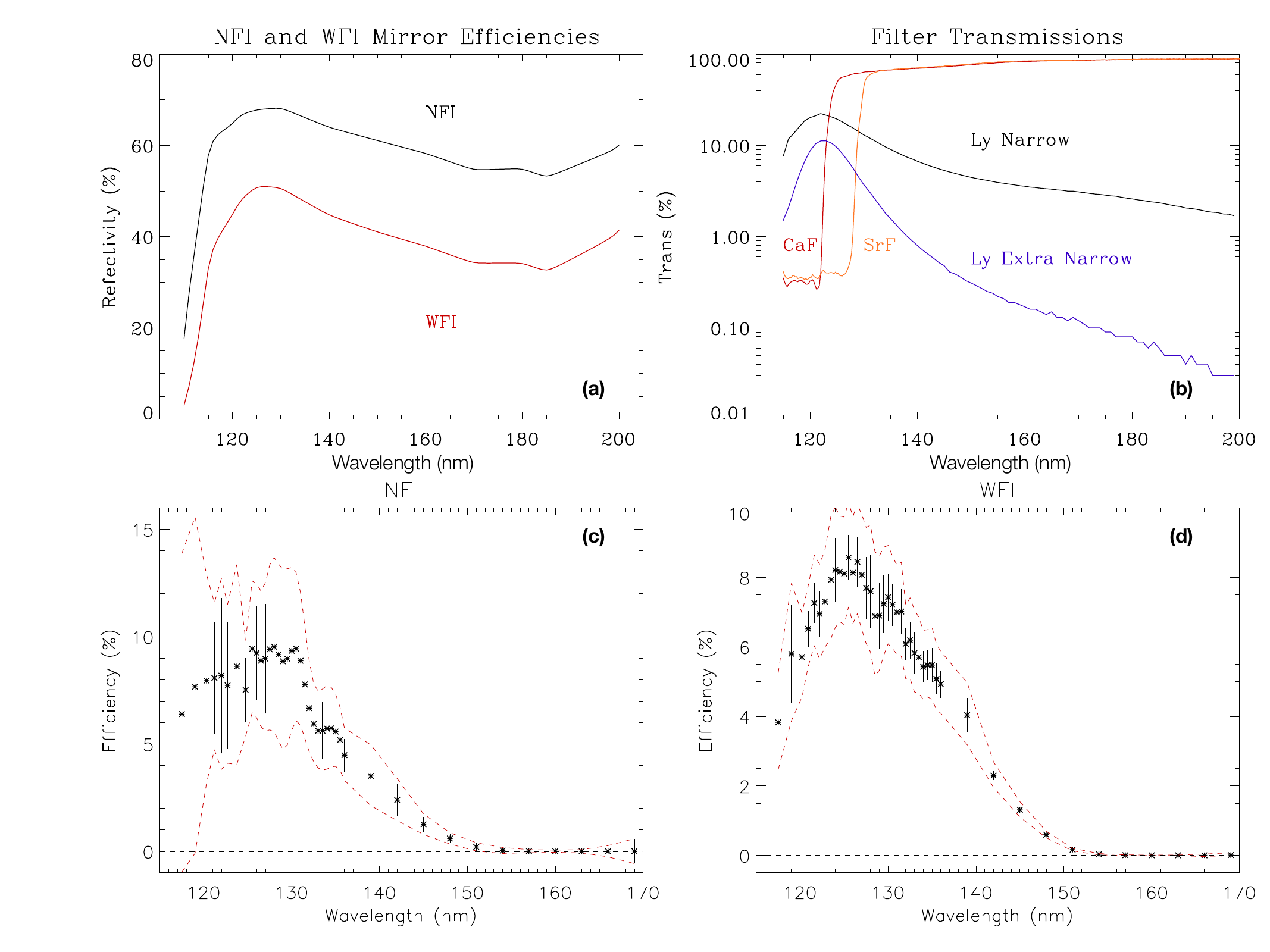}
\caption{Component efficiencies: (a) Al MgF$_2$ integrated mirror reflectivities, (b) Filter transmissions. Instrument efficiencies (Open filter position):(c) NFI, and (d) WFI.  Black bars denote random uncertainties. Red dashed lines include the systematic PMT calibration uncertainty of 10\% \cite{PTB25}. \color{black} (The large random uncertainties in efficiency of the NFI shortward of 125 nm (panel c) are due to low flux of the deuterium lamp source caused by a tarnished pick-off mirror; WFI measurements were made without this mirror.)\color{black}\label{fig:eff}}
\end{figure}

Input photon irradiance is measured contemporaneously to the detected spot counting rate measurements by means of a PTB-calibrated \cite{PTB25} photomultiplier tube (PMT) collecting light from a beam splitter in the optical GSE (OGSE).  The ratio of the irradiance at the GCI aperture to that at the monitor PMT is derived by using a second PTB-calibrated PMT to sample the beam incident on the GCI while simultaneously measuring with the monitor PMT.  Using this ratio, after subtraction of background currents, the monitor PMT photocurrent is converted to photon irradiance at the GCI aperture. The pupil diameter of the telescope is then applied to determine the total photon flux collected by the system. Dividing this into the detected counting rate yields the measured system efficiency as shown in Figure \ref{fig:eff}(c and d).  The absolute radiometric error-weighted throughput averaged over 120 to 125 nm for the WFI and NFI is 7.1\% $\pm$ 1.1 and 7.8\% $\pm$ 1.3, respectively.  On-orbit observations of stellar sources \color{black} bright in Lyman-$\alpha$ emission will be performed three times per year \color{black}
to monitor absolute radiometric response \cite{Zhang26c}(this issue).
%%%\color{black} Martin:  Talk about MCP response to low, med, and high flux here?  Linear, non-linear, and permanent damage?  Safety thresholds? Operational avoidance of bright stars? \color{black}

%%% Effectve areas from 120 to 125 nm are 0.383, and 0.0295 cm^2 for NFI, and WFI respt.

\begin{figure}   %%%  [<placement-specifier>]
\centering
\includegraphics[width=1.0\textwidth]{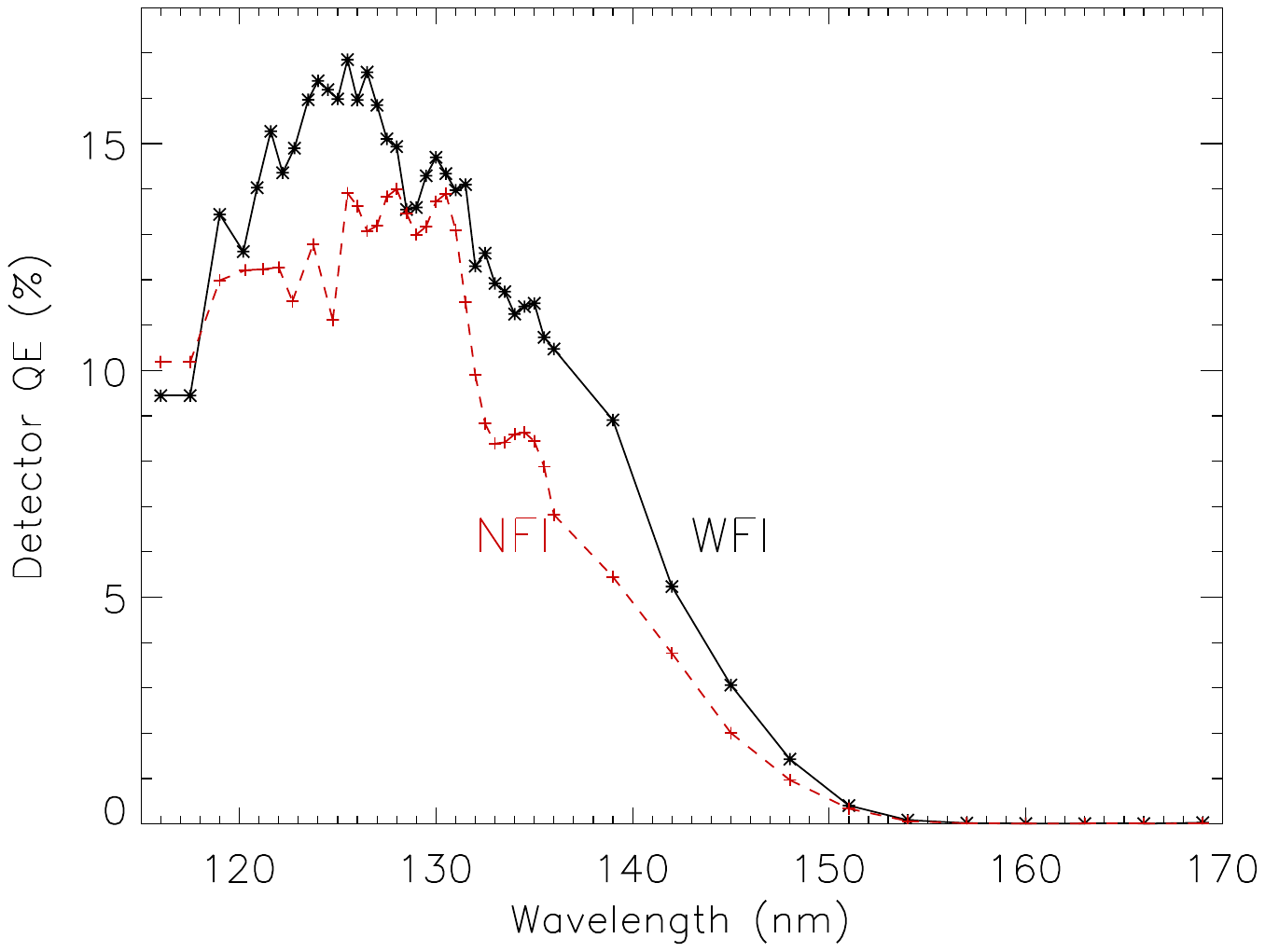}
\caption{Detector Quantum Efficiencies back-derived by dividing the measured Open position system efficiencies
(Figures \ref{fig:eff} c and d) by the measured mirror reflectivities (Figure \ref{fig:eff} a).}\label{detqe}
\end{figure}

The MCP detector QE is the product of the MgF$_2$ window, its 95\% transmissive Ni coating, and the QE of the KBr photocathode. The detector QEs were not measured directly; however, they were back-derived from the measured system sensitivities and measured integrated mirror reflectivity data.  The individual mirror reflectivities were acquired through witness sample measurements from Acton and GSFC before the mirrors were installed. Dividing these out from the measured system efficiencies found above yields the individual detector QEs shown in Figure \ref{detqe} (with uncertainties of the same relative proportion
as in Figure \ref{fig:eff}(c) and (d)).

\color{black}
Because the UV passband of operation the GCI is highly contamination sensitive, particularly to molecular contamination on the optics.  To maintain attention on this throughout the ground processing of the instrument, a Project-level risk was maintained from inception through launch.  To mitigate this risk the Project followed a stringent contamination control plan (CCP) (based on the previously implemented ICON-FUV CCP), implemented a T0 GN2 purge (which was started with initial optics installation into the instrument and maintained with minimal interruption through to launch). The optical cavity was kept closed except during specific testing under controlled conditions (the doors to the GCI imager were opened on-orbit well after the initial observatory outgassing period), testing of GCI throughput before and after instrument environmental testing, and the use of optical witness samples (OWS) in the GCI cavities (the reflectance of these OWS over the GCI passband were measured throughout Integration and Testing (I\&T), with particular attention around contamination-risky tests [TVAC, Vibration]).  The measurements from the OWS monitoring program are consistent with no throughput loss from contamination during I\&T.  The initial on-orbit throughput values for the GCI are consistent with the CSL lab measured throughputs to within 10\%.

Based on the gain at which the UV intensifiers are being operated and the expected counting rates, it is estimated that the MCP gain sag with mission extracted charge will not result in more than 1\% degradation of the throughput of the GCI. However, exposure of the MCPs to very bright sources will cause them to operate in a non-linear regime (temporary gain sag), or cause permanent gain loss.  The sun sensors automatically
turn off the detector HV should the Sun ever enter their FOV (Figure \ref{fig:fbd}).
Daily observations will avoid periods when very bright stars are within the FOV near Earth.
\color{black}

\subsection{Filter Transmissivities, Offsets,  and Thermal Drift}\label{filter_trans}

\begin{figure}   %%%  [<placement-specifier>]
\centering
\includegraphics[width=1.03\textwidth]{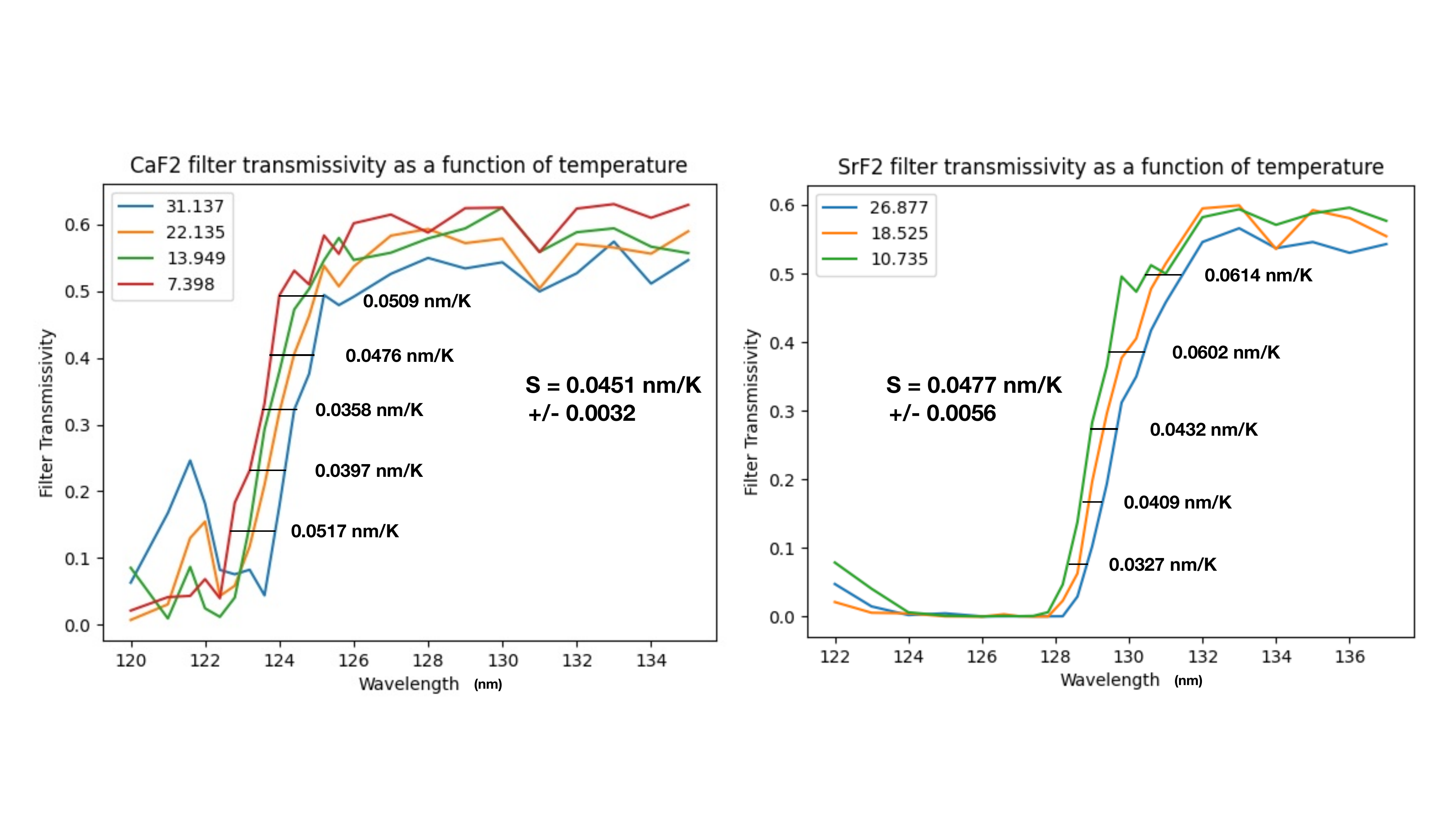}
\caption{NFI Filter edge wavelength shifts \emph{vs.} temperature in nm/K: (left) CaF$_2$ ; (right) SrF$_2$.  Insets indicate the temperature in $^\circ$C.} \label{fig:shifts}
\end{figure}

\begin{table}[h]
\caption{Measured Open Position and Filter Efficiencies in \% at Key Wavelengths (nm).}\label{tab:filt_eff}
\begin{tabular}{@{}llllrr@{}}
\toprule
Inst / Filter & 121.6 nm &    130.4 nm&    135.6 nm  & 150 nm & 170$^2$ nm \\
\hline
NFI Open    & 8.13     &    9.44  &    5.19 &  0.34 & $<$ 0.02\\
$\pm^1$     & 2.19     &    2.50  &    0.94 &  0.15 &          \\
\hline
NFI Ly-$\alpha$ Narrow &    1.178 &    0.692 &    0.293 & 0.03 & $<$ 0.02 \\
$\pm$                  &    0.256 &    0.157 &    0.078 & & \\
\hline
NFI Ly-$\alpha$  X-Narrow &    0.684 &    0.113 &    0.022 & $<$ 0.01 & $<$ 0.01 \\
$\pm$                     &    0.279 &    0.079 &    0.053 & & \\
\hline
  NFI CaF$_2$ &    0.018 &    4.972 &    3.271 & 0.26 & $<$ 0.02\\
$\pm$         &    0.157 &    0.898 &    0.479 & 0.15 &\\
\hline
  NFI SrF$_2$ &    0.017 &    5.146 &    3.712 & 0.26 & $<$ 0.02\\
$\pm$         &    0.193 &    0.989 &    0.584 & 0.15 & \\
\hline
&&& \\
&&& \\
\hline
WFI Open    &    7.265 &  7.21      &  5.08 & 0.33 & $<$ 0.02\\
$\pm$       &    0.574 &  0.607     & 0.436 & 0.10  & \\
\hline
WFI Ly-$\alpha$  Narrow &    1.075 &    0.557 &    0.235 & 0.025 & $<$ 0.01 \\
              $\pm$     &    0.111 &    0.045 &    0.027 & & \\
\hline
WFI Ly-$\alpha$  X-Narrow &    0.476 &    0.099 &    0.028 & $<$ 0.01 & $<$ 0.01\\
               $\pm$      &    0.096 &    0.037 &    0.023 & & \\
\hline
WFI CaF$_2$ &    0.010 &    4.060 &    2.798 & 0.24 & $<$0.02 \\
$\pm$       &    0.089 &    0.176 &    0.098 & 0.15 & \\
\hline
WFI SrF$_2$ &    0.012 &    3.484 &    2.738 & 0.24 & $<$0.01 \\
  $\pm$     &    0.098 &    0.147 &    0.096 & 0.15 & \\
\botrule
\end{tabular}
%% \footnotetext{Source: This is an example of table footnote.}
\footnotetext[1]{One $\sigma$ uncertainties}
\footnotetext[2]{Two $\sigma$ upper limit efficiency}

\end{table}
In addition to an Open and Closed position, each imager employs four filters: two narrow-pass centered on Lyman-$\alpha$ (made by Acton Optics),
and two long-pass windows with cut-off edges longward of Lyman-$\alpha$ (made by Crystran, measured by McPherson). The filter transmissions were measured as part of the system calibration in the vacuum chamber at CSL by determining relative throughput compared to the Open positions.  For each filter counting rate data was collected at filter-relevant wavelengths with the filter in place and with the Open position -- the ratio of these counting rates determines the in-system transmission of the filter under test. The narrow-pass filters (named Ly-$\alpha$ Narrow, and Ly-$\alpha$ Extra Narrow) show transmission widths (measured at FWHM) of 16.5, and 10.1 nm, respectively.  The long-pass filters both show a decrease in transmission
by a factor of 10 for a 1.7 nm decrease in wavelength (see Figures \ref{fig:eff} and \ref{fig:shifts}).  The system throughput as observed in the Open position and through the four filters is presented in Table \ref{tab:filt_eff} for the key wavelengths
Lyman-$\alpha$ 121.6 nm, OI 130.4 and 135.6 nm, and at 150 and 170 nm.
Longward of 155 nm the Open filter position shows $<$~0.1\% efficiency, hence,
the filters show even lower efficiency (shown in Figure \ref{fig:filter_trans}).
No significant signal was detected at 170 nm in either channel; therefore, the efficiencies at
this wavelength are presented as 2$\sigma$ upper limits in Table \ref{tab:filt_eff}.
Changes in sensitivity as a function of angle-of-incidence are less than 1\%.

Crystalline fluorides are known to show filter edge drifts as a function of temperature
%%% (Laufer, 1965)
\cite{Laufer65}.  The rate of thermal drift of the filter edge is important to characterize on the ground so that measurements of terrestrial O and N$_2$ emissions on-orbit using these filters are accurate.  Thermal drift of the filter edge was measured by performing filter transmission measurements for the long-pass filters with a fine sampling grid of 0.2 nm at the transmission cut-off edge while the temperature of the GCI was slowly varied from 35 to 5 $^\circ$C. The measured thermal drift of the NFI filter edges for SrF$_2$ and CaF$_2$ is $+0.0477 \pm 0.0056$, and $+0.0451 \pm 0.0032$  nm/K, respectively, as shown in Figure \ref{fig:shifts}. 
%% Double check these numbers !!!  Done MMS
Our value for CaF$_2$ is 13\% ($1.1 \sigma$) greater than 0.040 nm/K found by
%%% Laufer et al., (1965) 
\cite{Laufer65}. \color{black}To date, the observed range of on-orbit temperature variations during nadir pointings has been 0.7 $^\circ$C. Observations at angles 41 degrees from the anti-Sun direction
have shown an additional 0.5 $^\circ$C variation.
%%  Are these values consistent with the values in the literature for these crystals? Yes, within 1.1 sigma for CaF2 ! MMS
\color{black}

Because of their refractive properties, the crystalline CaF$_2$ and SrF$_2$ windows can introduce a small deviation in beam direction.
During the initial vacuum UV checkout tests at CSL we measured an $\sim2$ pixel shift in the measured spot centroid of the NFI between
the Open position and the two crystalline filters.  This is a significant fraction of a resolution element and a more detailed calibration of this shift will be performed on-orbit for both imagers by observing stars at different filter positions \cite{Waldrop26}(this issue).

\subsection{Background}\label{subsec5.7}

The MCP and CMOS elements of the detectors are each sources of random background noise.
On the ground the intrinsic background rate of the MCPs is very low, whereas that of the CMOS detectors is about 60 times greater (see Table \ref{tab:mcpcmos}).
On-orbit the major source of background will be solar energetic particles (SEP). 
Since MCPs are about 100 times more sensitive to energetic particles than CMOS detectors, they will dominate the background signal.
\color{black}The effects of SEP and cosmic rays on the MCPs are mitigated by the averaging of many CMOS frames (480 per minute) for the nominal 30 and 60 minute science integrations. \color{black}
However, knowing the locations of noisy pixels of the CMOS detectors is necessary
to prevent them from contaminating real signal.
%%; $\sim$ 1 count/s/cm$^2$, whereas that of the CMOS detectors is much greater; $\sim 5 \times 10^{6}$ e$^-$/s/cm$^2$.
Several long-duration background images were obtained at CSL for both the NFI and WFI.  Because no corresponding bias frames for these images were obtained at CSL \color{black} (because of time constraints)\color{black}, all data were processed
by subtracting column-median values.  Thus, dark frames show a distribution of
count values centered at zero counts --- absolute background count rates cannot
be determined directly, but can be estimated by assuming they follow a Poisson distribution.
The dark frame count-per-pixel histograms show three distinct populations: normal, warm, and hot.  The normal pixel population comprises $\sim$ 99\% of the pixels and shows
about 0.2 and 0.4 count/pixel/min for the NFI and WFI imagers, respectively.
About 1\% of pixels show count rates 6 times greater than the standard deviation of the normal-pixel distribution and
will need attention: either via flat-field division for warm (and cool) pixels,
or local-median replacement for hot pixels. \color{black}This will be performed
on the ground as part of the standard science pipeline processing. \color{black}
Dark images obtained in the Closed position, and flat-field images acquired on-orbit 
(obtained by imaging the IPH far off-nadir in multiple spacecraft orientations) 
will track both temporal and thermal variations in individual pixel sensitivity and noise levels %%%(Waldrop et al., 2025, this issue)
\cite{Waldrop26,Zhang26a,Zhang26b,Zhang26c}(this issue).

\subsection{Stray Light}\label{subsec5.8}

Stray light (sneak path) tests were performed
at SSL by shining white light into one channel
while checking for signal in the other channel (with its filter wheel in the Closed position).
No light-leaks were seen, but, the degree of suppression was not quantified. No stray light was noticed in either channel in any of the UV images obtained at CSL, however, cross-talk between channels was not explicitly checked at UV wavelengths.

\noindent
%For a quote environment, use \verb+\begin{quote}...\end{quote}+
%\begin{quote}
%Quoted text example. Aliquam porttitor quam a lacus. Praesent vel arcu ut tortor cursus volutpat. In vitae pede quis diam bibendum placerat. Fusce elementum
%convallis neque. Sed dolor orci, scelerisque ac, dapibus nec, ultricies ut, mi. Duis nec dui quis leo %sagittis commodo.
%\end{quote}

%\section{Methods}\label{secnn}

%Topical subheadings are allowed. Authors must ensure that their Methods section includes adequate experimental and characterization data necessary for others in the field to reproduce their work. Authors are encouraged to include RIIDs where appropriate. 

%\textbf{Ethical approval declarations} (only required where applicable) Any article reporting experiment/s carried out on (i)~live vertebrate (or higher invertebrates), (ii)~humans or (iii)~human samples must include an unambiguous statement within the methods section that meets the following requirements: 

%\begin{enumerate}[1.]
%\item Approval: a statement which confirms that all experimental protocols were approved by a named institutional and/or licensing committee. Please identify the approving body in the methods section

%\item Accordance: a statement explicitly saying that the methods were carried out in accordance with the relevant guidelines and regulations
%\end{enumerate}[1.]

\section{\color{black}Summary}\label{sec6}

In Section \ref{intro} we outline the over-arching goals of the Carruthers mission
which are to understand the dynamics of the Earth's exosphere.
Because the exosphere is predominantly composed of Hydrogen the study of its emissions can
yield its distribution and density. Solar radiation (high energy particles (SEP) and
EUV light) both cause H atoms to emit in the EUV at 121.6 nm (Lyman-$\alpha$).
The study of this emission at moderately high spatial and temporal resolutions
will provide insight into the physical processes that give the exosphere its form.
The study of the exosphere is best done from a vantage point completely outside of it.
A halo orbit around the Sun-Earth Lagrange L1 point provides a continuous view of the Earth while the solar panels provide continuous power and full shade to the instrument entrance apertures.
In Section \ref{sec2} the specific science requirements are outlined which then directly
drive the payload and instrument requirements described in Section \ref{sec3}.

Based on the goals and requirements stated above, an instrument design was created with
sufficient margin in the various components to accommodate any deficiencies or
changes in the subsequent design.  The overall strategy is to provide adequate spatial
resolution near the exobase, while simultaneously observing the much fainter outer exosphere.
This is accomplished by employing co-aligned narrow- and wide-field imagers.

Each imager is equipped with a filter set designed to
maximize the contrast
between in- and out-of- band emissions,
\color{black} as well as provide the high dynamic range required to image the inner and outer
regions of the exosphere.
\color{black}
The overall passband of the instrument is governed by the detectors' MgF$_2$ windows
which cut off below 115 nm, and their KBr photocathodes which are insensitive
longward of 155 nm, and the telescope mirror reflectivities which, by design, peak around 125 nm.
To distinguish between Lyman-$\alpha$ at 121.6 nm
and O I at 130.4 and 135.6 nm, we employ two narrow-pass filters centered at 121.6 nm,
and two long-pass windows that discriminate against Lyman-$\alpha$ but allow the passage of longer wavelengths.

%%% 

The telescopes were first aligned at visual wavelengths and then tested in the vacuum ultraviolet.
The various tests performed have directly measured or allowed the determination of the following parameters:
spot size and shape (PSF), image scale, field of view, geometric distortion, relative filter transmissions,
and filter-edge thermal drifts.
The transmissions from the narrow-pass filters and the long-pass windows both show the
contrast required to segregate Lyman-$\alpha$ from OI  and N$_2$ emissions.
The absolute radiometric throughput (in Open filter position) for both imagers is quantified to within 20\%.
Ground-based background count rates are very small, but the CMOS detectors show $\sim$ 1\% warm/hot pixels.
No light leaks were observed either at visual or UV wavelengths.
The specific parameter values are presented in the Tables and Figures
of Section \ref{calibrations}.

%%% (see Figure \ref{fig:filter_trans}, and Table \ref{tab:filt_eff}).
%%More thoughts .....  Encircled Energy (at 95\ and 98\%) is quantified. 

The observed PSFs show Lorentzian wings, but they are not problematic.
Deconvolution techniques (e.g. maximum entropy \cite{Hollis92}) applied to synthetic data convolved with the observed NFI PSF are able to recover
sharp gradients in the inner exosphere to within 1\% accuracy.

\color{black}
On-orbit calibrations are necessary to verify and/or refine all calibration results of this paper. Three Ly-$\alpha$-bright stars (Capella, $\alpha$ Boo, and HR1099) will be observed periodically to track absolute throughput.
Observations of star fields near the galactic plane will characterize
the PSFs, in-band scattering levels, positional filter shifts, and optical distortions.  Observations of the IPH far from nadir in multiple spacecraft orientations will provide detector flat fields, track noise levels of
individual pixels, and map the intensity of the IPH.  Targets of opportunity (e.g. Jupiter, Saturn, comets)
will provide potentially useful data for both calibration and scientific purposes.
\color{black}

%%% Bullet list of intended on-orbit calibrations ???  Absolute eff.  PSF,  Scattering, Distortions, IPH, Flat Fields, Targets of Oportunity ?
%%%% Preliminary analysis of in-flight images of several stars show PSF widths consistent with ground-based measurements.

The instrument meets all performance requirements, most with large margin.
Calibration files are provided as additional on-line content \cite{sirk26b} which will be useful for GCI researchers as well as future UV instrument designers.

%Discussions should be brief and focused. In some disciplines use of Discussion or `Conclusion' is interchangeable. It is not mandatory to use both. Some journals prefer a section `Results and Discussion' followed by a section `Conclusion'. Please refer to Journal-level guidance for any specific requirements. 

%\section{Conclusion}\label{sec13}

%Conclusions may be used to restate your hypothesis or research question, restate your major findings, explain the relevance and the added value of your work, highlight any limitations of your study, describe future directions for research and recommendations.In some disciplines use of Discussion or 'Conclusion' is interchangeable. It is not mandatory to use both. Please refer to Journal-level guidance for any specific requirements. 

\backmatter

%\bmhead{Supplementary information}
%If your article has accompanying supplementary file/s please state so here. 

%Authors reporting data from electrophoretic gels and blots should supply the full unprocessed scans for key as part of their Supplementary information. This may be requested by the editorial team/s if it is missing.

%Please refer to Journal-level guidance for any specific requirements.

\bmhead{Acknowledgments}

%Acknowledgements are not compulsory. Where included they should be brief. Grant or contribution numbers may be acknowledged.

%Please refer to Journal-level guidance for any specific requirements.

%\section*{Declarations}

%Some journals require declarations to be submitted in a standardised format. Please check the Instructions for Authors of the journal to which you are submitting to see if you need to complete this section. If yes, your manuscript must contain the following sections under the heading `Declarations':

\begin{itemize}
\item Funding: This work was supported by the NASA Science Mission Directorate, Heliophysics Division through contract 80GSFC21C0038. Additional support was provided by the Belgian Federal Science Office.
\item Conflict of interest/Competing interests:
Not applicable
\item Ethics approval and consent to participate:
Not applicable
\item Consent for publication:
Not applicable
\item Data availability:
https://doi.org/10.5281/zenodo.16862116 \cite{sirk26b}
\item Materials availability:
Not applicable
\item Code availability: Included with Data \cite{sirk26b}
\item Author contribution:
Ordered in author list.
\end{itemize}

\addcontentsline{toc}{section}{References}
\printbibliography %Prints bibliography

\end{document}